\documentclass{article}
\usepackage[dvipsnames]{xcolor} 
\usepackage{adjustbox}
\usepackage[ruled]{algorithm2e}
\usepackage{amsfonts}
\usepackage{amsmath}
\usepackage{amsthm}
\usepackage{authblk}
\usepackage[backend=biber,citestyle=authoryear-comp, sorting = nyt, maxcitenames=3, mincitenames=1, maxbibnames = 20]{biblatex}
\DeclareSourcemap{
  \maps[datatype=biblatex]{
    \map{
      \step[fieldsource=pmid, fieldtarget=usera]
      \step[fieldsource=pmcid, fieldtarget=userb]
    }
  }
}

\DeclareFieldFormat{usera}{PMID: #1}
\DeclareFieldFormat{userb}{PMCID: #1}

\renewbibmacro*{finentry}{%
  \setunit{\adddot\space}%
  \printfield{usera}%
  \newunit\newblock
  \printfield{userb}%
  \finentry}
  
\usepackage{bigints}
\usepackage{bm}
\usepackage{booktabs}
\usepackage{ctable}
\usepackage{float}
\usepackage{graphicx}
\usepackage{geometry}
\usepackage{makecell} 
\usepackage{mathtools}
\usepackage{setspace}
\usepackage{siunitx}
\usepackage{subcaption}

\usepackage{hyperref} 

\hypersetup{
    colorlinks=true,
    citecolor = blue,
    linkcolor=blue,
    filecolor=magenta,      
    urlcolor=cyan,
    pdfpagemode=FullScreen,
    }
\makeatletter
\renewcommand\AB@affilsepx{ \quad } 
\makeatother
\DeclareCiteCommand{\citeyearbracket}
  {\usebibmacro{prenote}}
  {\printnames{labelname}\space(\printfield{year})}
  {\multicitedelim}
  {\usebibmacro{postnote}}

\newcommand{\allpi}{\bm{\pi}}

\newcommand{\half}{1/2} 
\newcommand{\bg}{\beta_g}
\newcommand{\bDeltag}{\bm{\Delta}_g }

\newcommand{\bGammacg}{\bm{\gamma}_{c,g}}
\newcommand{\blamg}{\bm{\lambda}_g }
\newcommand{\ku}{ \kappa }

\newcommand{\bmug}{\bm{\mu}_g }

\newcommand{\bphip}{\bm{\phi}_p}
\newcommand{\bSigg}{\bm{\Sigma}_g}

\newcommand{\bX}{\bm{X}}
\newcommand{\bXo}{\bm{X}^{o}}
\newcommand{\bXm}{\bm{X}^{m}}
\newcommand{\bxo}{\bm{x}^{o}}
\newcommand{\bxm}{\bm{x}^{m}}
\newcommand{\bx}{\bm{x}}

\newcommand{\bmu}{ \bm{\mu}}
\newcommand{\bSig}{ \bm{\Sigma}}
\newcommand{\blam}{ \bm{\lambda}}

\newcommand{\bDelta}{ \bm{\Delta}}
\newcommand{\btheta}{ \bm{\theta}}
\newcommand{\bOmega}{ \bm{\Omega}}
\newcommand{\bOmegag}{ \bm{\Omega}_g}
\newcommand{\bZ}{ \bm{Z}}
\newcommand{\bz}{ \bm{z}}
\newcommand{\cD}{ \mathcal{D}}

\newcommand\deq{\stackrel{d}{=}}

\newcommand{\et}{ t^{(k)}_{ig}}

\newcommand{\etwo}{t^{2 (k)}_{ig} }
\newcommand{\ev}{ \mathbb{E}}

\newcommand{\hag}{\widehat{\alpha}}
\newcommand{\hbg}{\widehat{\beta}}

\newcommand{\heta}{\eta^{(k)}_{ig}}
\newcommand{\hetac}{\eta^{(k)}_{\beta, ig}}
\newcommand{\hlg}{\widehat{\bm{\lambda}}}
\newcommand{\hmg}{\widehat{\bm{\mu}}}

\newcommand{\hsg}{\widehat{\bm{\Sigma}}}
\newcommand{\uk}{\textbf{u}^{(k)}_{ig} }
\newcommand{\uck}{\textbf{u}^{c(k)}_{ig} }
\newcommand{\vk}{v^{(k)}_{ig} }
\newcommand{\vctxk}{\Tilde{E}^{(k)}_{vt\bm{x},ig} }
\newcommand{\vHk}{\textbf{H}_{ig}^{(k)}}
\newcommand{\vhk}{\textbf{h}_{ig}^{(k)}}
\newcommand{\vHck}{\textbf{H}_{ig}^{c(k)}}
\newcommand{\vhck}{\textbf{h}_{ig}^{c(k)}}
\newcommand{\vkt}{vt^{(k)}_{ig} }
\newcommand{\vktx}{E^{(k)}_{vt\bm{x},ig} }
\newcommand{\vktt}{vt^{2 (k)}_{ig} }
\newcommand{\vx}{E^{(k)}_{v\bm{x},ig} }
\newcommand{\vcxx}{\Tilde{E}^{(k)}_{v\bm{x}\bm{x}^{\top},ig} }
\newcommand{\vxx}{E^{(k)}_{v\bm{x}\bm{x}^{\top},ig} }
\newcommand{\vcxk}{\Tilde{E}^{(k)}_{v\bm{x},ig} }
\newcommand{\zk}{z^{(k)}_{ig} }

\newtheorem{theorem}{Theorem}[section]

\title{Sleep pattern profiling using a finite mixture of contaminated multivariate skew-normal distributions on incomplete data}
\author[1]{Pillay J.}
\author[2]{Tortora C.}
\author[3]{Punzo A.}
\author[1]{Bekker A.}
\affil[1]{University of Pretoria} \affil[2]{San José State University} \affil[3]{University of Catania}

\date{}

\begin{document}

\maketitle

\begin{abstract}
Medical data often exhibit characteristics that make cluster analysis particularly challenging, such as missing values, outliers, and cluster features like skewness.
Typically, such data would need to be preprocessed---by cleaning outliers and missing values---before clustering could be performed. 
However, these preliminary steps rely on objective functions different from those used in the clustering stage.
In this paper, we propose a unified model-based clustering approach that simultaneously handles atypical observations, missing values, and cluster-wise skewness within a single framework. 
Each cluster is modelled using a contaminated multivariate skew-normal distribution—a convenient two-component mixture of multivariate skew-normal densities—in which one component represents the main data (the “bulk”) and the other captures potential outliers.
From an inferential perspective, we implement and use a variant of the EM algorithm to obtain the maximum likelihood estimates of the model parameters.
Simulation studies demonstrate that the proposed model outperforms existing approaches in both clustering accuracy and outlier detection, across low- and high-dimensional settings, even in the presence of substantial missingness. 
The method is further applied to the Cleveland Children’s Sleep and Health Study (CCSHS), a dataset characterised by incomplete observations. 
Without any preprocessing, the proposed approach identifies five distinct groups of sleepers, revealing meaningful differences in sleeper typologies.

\end{abstract}
{\bf Keywords:} augmented EM type, contaminated mixture models, missing values at random, outliers, skew-normal distribution.

\section{Introduction}
\label{introduction}

Sleep data is often multivariate, continuous, and characterised by complex cluster structures, skewness and leptokurtosis (\cite{agewise}). This cluster-wise skewness is particularly prominent in sleep research, where patient-level variability, and irregular sleep patterns naturally give rise to asymmetry in the clusters of individuals or patients (\cite{agewise, slp_fmm, slp_glc_skew, slp_actigraphy}). Finite mixture models are a valuable tool to describe and interpret the heterogeneity in the sleeping patterns (\cite{slpapnea_kmeans, slp_gmm, slp_gmm_spindle, slp_ica}). Specifically, finite mixtures of skew normal distributions are not only theoretically convenient, but computationally attractive to fit to sleep data. Noticeably, when cluster distributions are not symmetric and have heavier tails, symmetric distributions often compensate by overestimating the number of clusters. The 'extra' clusters are not helpful in practice and weaken interpretations and analyses (\cite{slp_cls_num, cls_luca, cls_mclachlan}).
\\
\\
Moreover, incomplete patient information is a recurring challenge. There are numerous sleep studies that have had to apply preprocessing techniques to the raw datasets because they were incomplete, including \cite{missing_apnea1, missing_phenotyping, missing_hierarchical} and \cite{missing_ehlers_danlos} to name a few. There are also systematic sleep study reviews that collate raw data: these reviews experience the worse end of missing values as a problem - see \cite{missing_polysomnography} and \cite{missing_diabetes} for more details. Missing values may arise from numerous sources such as technical failures in recording devices, patient non-compliance, incomplete clinical records, or incomplete patient self-reporting. In statistical analysis, an assumption must be made about the relationship between the probability of data being missing and the underlying values of the variables involved in the analysis. The statistical modelling of missing values can be classified into three broad categories (\cite{missingess}), namely: missing not at random (MNAR), missing at random (MAR), and missing completely at random (MCAR). MNAR data assume the probability of missingness patterns in the data depends on the value of the missing entries. MAR data, however, assume the patterns behind the missing values do not depend on the value of the missing entries, but may depend on the value of the observed components of the observed vector. Lastly, MCAR data do not assume dependence on any values of the data. MAR is a consequence of several factors outside the control of the data collection process. The patterns in missingness could therefore be completely random, or may be linked to the observed points in the same observed vector. The latter mechanism is more plausible and common in sleep datasets. Furthermore, the definitions of MCAR and MAR imply that MCAR is a special case of MAR. Traditional pre-processing techniques typically address these issues in one of two ways (\cite{missing_handling}): deletion or multiple imputation. The former excludes meaningful patient-related information, weakening the representative power of the data with respect to its patients' backgrounds (\cite{apnea_time_series}). The latter occurs in two separate stages. First, missing values are imputed using methods separate from the clustering model that do not account for the clustering structure. Only afterwards, clustering is performed on the imputed data. This sequential approach may inadvertently distort the underlying structure by ignoring the interplay between missingness and skewness affecting each cluster, leading to biased parameter estimates and weaker interpretability. To overcome these limitations, this paper proposes an algorithm that performs simultaneous clustering and imputation of missing values. The advantage of this approach lies in the fact that the same statistical model that best captures the data’s structure is also used to impute the missing values, ensuring consistency between the two tasks. \\
\\

In addition to the presence of missingness, sleep data also present other challenges, such as atypical points that deviate from the apparent groupings. These points are referred to often as outliers, and it is of equal, if not higher, importance to understand the reasons for their deviation in the data, as they can represent rare, but meaningful diagnostic relevance in people's lives (\cite{medical}). People are unique and have innumerable responses to stimuli, which makes outliers a common issue. However, the term outliers conceals meaning beyond explainable values. From a statistical perspective, an outlier can be categorised as one of two types (\cite{outliers}): Mild outliers are considered far from the population's distribution or even following a different distribution. These points can be identified, predicted, and explained sufficiently by distribution-based models. Gross outliers, on the other hand, have no pattern to them, and no probability distribution can sufficiently model these points. Gross outliers are unpredictable. It is left to the analyst to trim the outliers according to ad hoc measures. The model proposed in this paper naturally accommodates mild outlier detection, flagging data points that deviate substantially from cluster-specific distributions.\\

This paper addresses the problems mentioned above by considering a finite mixture of contaminated skew-normal distributions (FMCMSN) on incomplete data. The finite mixture addresses heterogeneity in the data by describing the driving force behind said heterogeneity as an additive mix of homogeneous clusters. The multivariate skew-normal (MSN) distribution is a strong candidate that can fit and explain asymmetry within each cluster. A mixture model of contaminated MSN distributions can detect potential mild outliers within each clusters. It extends the concept of a contaminated skew normal distribution from \cite{lachos2010likelihood}. It is also possible to do outlier detection under this concept of contamination. The paper proposes an EM type algorithm that is altered to simultaneously handle missing values under the MAR mechanism to fit the FMCMSN model to incomplete datasets. The result is a set of estimated parameters that explain the features of each cluster, the proportion of outliers in each cluster, and a technique to classify each point as an outlier through a posteriori probabilities. The rest of this paper is structured as follows: Section \ref{background} introduces the skew-normal distribution and its contaminated formulation. Section \ref{estimation} discusses the clustering capabilities of the model and its parameter estimation under the MAR mechanism. These are, however, symmetric distributions that do not account for cluster-dependent asymmetry. Thus, Section \ref{simulation} conducts an extensive simulation study of the FMCMSN's performance and its comparison with competitors. Section \ref{application} demonstrates the application of the FMCMSN model to the Cleveland sleep study (incomplete) dataset, which provides five clusters, each with a unique sleeping pattern. The results also identify atypical observations across two clusters in the dataset, allowing for further interpretation of the causes for these observations. Section \ref{conclusion} concludes with areas for future work.

\section{Background}
\label{background}
Model-based clustering is an approach that describes the grouping structure in data as a finite mixture model, each homogeneous group is governed by its own probability distribution. Formally, an observed vector $\bx \in \mathbb{R}^p$ comes from a $G$-component finite mixture model with probability density function (pdf):

\begin{align}
    \label{fmm}
    f(\bx; \bm{\psi}) = \sum_{i=1}^G\pi_gf(\bx;\btheta_g),
\end{align}
where $f(\bx;\theta_g)$ denotes the pdf of the $g^{th}$ component and its corresponding set of parameters $\btheta_g$ and $\bm{\psi} = \{ \btheta_g\}_{g=1}^G$ . The mixing probability $\pi_g$ for the $g^{th}$ cluster acts as a weight and is subjected to the constraint $\pi_g>0$ and $\sum_{g=1}^G \pi_g = 1$. Here, $\bm{\psi}$ denotes the collection of all parameters in the mixture model. The formulation in \eqref{fmm} is linear and affords attractive analytic and computational tractability. When the component pdfs are normal with mean $\bmug$ and covariance matrix $\bSigg$, meaning $\bX|g \sim N(\bmug, \bSigg)$, the model is popularly known as a Gaussian Mixture Model (GMM). There is extensive use on GMMs in literature due to its algebraic ease, but they assume that each cluster's distribution is elliptically symmetric. As introduced in Section \ref{introduction}, there are datasets that do not adhere to this assumption due to their inherent skewness. 

\subsection{The multivariate skew-normal distribution}
The multivariate skew normal distribution, introduced by \cite{azzalini1996}, extends the multivariate normal (MN) distribution so it can handle asymmetric data. Formally, an observation $\bx \in \mathbb{R}^p$ that is generated by a MSN distribution has the following pdf:
\begin{align}
\label{skew normal pdf}
    f_{MSN}(\bx;\bmu,\bSig,\blam, \lambda_0) = \frac{1 }{ \Phi_1\left( \frac{\lambda_0}{\sqrt{1 + \bm{\blam}^{\top}\bm{\blam} } } \right) }\bphip(\bx;\bmu, \bSig) \Phi_1\left( \lambda_0 + \blam^{\top}\bSig^{-\half}(\bx - \bmu)  \right),
\end{align}
where $\bm{\phi}_p(\cdot;\bmu,\bSig)$ is the pdf of a  MN distribution with mean $\bmu$ and covariance matrix $\bSig$ and $\Phi_1(\cdot)$ denotes the univariate standard normal cumulative distribution function (cdf). Parameters $\bmu \in \mathbb{R}^p$ and $\bSig \in \mathbb{R}^{p\times p}$ are the respective location vector and scale matrix of the distribution, subjected to the constraint that $\bSig$ is a positive definite matrix. The remaining parameters are the skewness vector $\blam \in \mathbb{R}^p$ and the threshold $\lambda_0 \in \mathbb{R}$. In \cite{hiddentruncationmodels}, the derivation of the MSN distribution begins by truncating elements of a normally distributed random vector against a threshold $\lambda_0$. The threshold parameter $\lambda_0$ arises in the construction of the MSN distribution through a method that involves truncating a bivariate normal distribution—specifically, by retaining only those observations where one component exceeds a certain threshold. From a simulation perspective, $\lambda_0$ represents this cutoff, effectively filtering the data to induce asymmetry, which directly contributes to the skewness of the resulting distribution. However, because skewness can also be controlled by another parameter, namely $\blam$, $\lambda_0$ is typically set to zero to simplify the model and avoid complex interactions between the two parameters. In this paper, the clusters are assumed to be generated by a MSN distribution with pdf given in \eqref{skew normal pdf}, but with $\lambda_0$ set to 0 to explore the behaviour and performance of $\blam$. The notation  $\bX \sim SN\left(\bmu, \bSig, \blam \right)$ is used to denote that $\bX$ follows a MSN distribution with $\lambda_0 = 0$. Setting $\blam = \bm{0}$ simplifies the pdf in \eqref{skew normal pdf} to a $N(\bmu, \bSig)$ distribution. That is, the MSN distribution does not exclude symmetry as a property in data, but includes it as a special case, making it a more flexible option to the MN distribution. 

\subsection{The contaminated multivariate skew-normal distribution }
To simultaneously handle mild outliers in data with an asymmetric distribution, a contaminated MSN model (CMSN) is revisited. Specifically, an observation $\bx \in \mathbb{R}^p$ is generated by a CMSN distribution with the following pdf:
\begin{align}
    \label{contaminated skew normal pdf}
    f_{CMSN}(\bx;\bmu,\bSig,\blam, \alpha, \beta) = \underbrace{\alpha f_{MSN}(\bx;\bmu,\bSig,\blam)}_{\text{good component}} + \underbrace{( 1 - \alpha)f_{MSN}(\bx;\bmu,\beta\bSig,\blam)}_{\text{bad component}},
\end{align}
where $\alpha$ may be interpreted as the proportion of typical data (hereinafter referred to as good points) and the inflation parameter $\beta >1$ may be interpreted as the degree of contamination. A random vector $\bX$ is said to follow the distribution with the pdf given in \eqref{contaminated skew normal pdf} using the notation $\bX \sim CMSN(\bmu,\bSig,\blam,\alpha,\beta)$. Notice that the distribution with pdf \eqref{contaminated skew normal pdf} is a two-component mixture form of the FMM in \eqref{fmm}, in which one component, with mixing probability $\alpha$, represents the good observations, and the other component, with mixing probability $1 - \alpha$, represents the anomalous points (or outliers) referred to as bad points. The location parameters for both components are the same, however the component that contains the anomalies has an inflated version of the population's scale matrix, given as $\beta \bSig$.  When $\beta \rightarrow 1$ and $\alpha \rightarrow 1$ the pdf in \eqref{contaminated skew normal pdf} reduces to an ordinary MSN pdf as introduced in \eqref{skew normal pdf}. That is, the CMSN model is able to model a dataset's distribution even if there are no bad points present. 

It is algebraically attractive to recognise that a random vector $\bX\sim CMSN(\bmu,\bSig,\blam,\alpha,\beta)$ has the following stochastic representation (\cite{lachos2007likelihood, lachos2010likelihood}):
        \begin{align}
            \bX \deq  \bmu +\sqrt{K} T \bm{\bDelta} +  \sqrt{K}\bSig^{1/2}(\bm{I} -\bm{\delta} \bm{\delta}^{\top} )^{1/2} \bm{Y},
            \label{sr}
        \end{align}
   where $\bDelta = \frac{\bSig^{1/2}\blam}{ \sqrt{ 1 + \blam^{\top}\blam}  } $, $K = V + (1-V)\beta$, $\bm{\delta} = \frac{\blam}{ \sqrt{ 1 + \blam^{\top}\blam}  } $, $\bm{Y} \sim N(\bm{0}, \bm{I})$ independent of univariate random variable $T$ which follows a truncated standard normal distribution (TN) on the interval $(0,\infty)$ denoted as $T\sim TN(0,1)$. The random variable $V$ follows a Bernoulli distribution with parameter $\alpha$, and dictates whether or not $\bX$ is generated by the good component of the CMSN model with probability $\alpha$.


\subsection{FMCMSN model for incomplete observations}
\label{fmcmsn for missing}
A finite mixture of contaminated multivariate skew-normal (FMCMSN) distributions follows from substituting \eqref{contaminated skew normal pdf} into \eqref{fmm} which produces the following pdf:
\begin{align}
\label{FMCMSN}
    f_{FMCMSN}(\bx; \bm{\psi}) = \sum_{g=1}^G \pi_gf_{CMSN}(\bx;\bmug,\bSigg,\blamg, \alpha_g,\bg).
\end{align}

The model in \eqref{FMCMSN} must be altered to account for observed vectors that are incomplete. The first step in this process is to denote the observed and missing components of a random vector. Thus, $\bX$ is decomposed into its missing and observed vectors $\bX=[ \bXm, \bXo]^{\top}$. The superscripts $o$ and $m$ indicate the observed and missing parts of $\bX$, respectively. The missing component of $\bX$ is assumed to be missing at random (MAR). Thus, the probability of what is missing does not depend on the value of the missing part itself, but may depend on, or be better informed by, the observed component. This realisation allows for some closed-form distributions of the missing components conditioned on the observed components. These closed-form results apply at the cluster level. The random variable $V$ in \eqref{sr} identifies whether $\bX$ is an outlier relative to the $g^{th}$ cluster. We therefore introduce $V_g$ to denote this variable for cluster $g$. It takes the value indicating outlier status with probability $1-\alpha_g$. The closed-form distributions for the missing component of $\bX$ appear in Theorem \ref{sn dists} and Theorem \ref{norm cond dist}.



   \begin{theorem}
   \label{sn dists}
   Consider $ \bX|g \sim CMSN(\bmug,\bSigg,\blamg,\alpha_g,\beta_g)$. Partition the random vector $\bX$ and its distribution parameters in terms of the $\bX$'s observed and missing components as:
        \begin{align}
        \bX = \begin{bmatrix} \bXm \\ \bXo \end{bmatrix},~
        \bmug = \begin{bmatrix} \bmu_{o,g} \\ \bmu_{m,g} \end{bmatrix},~ \bSigg = \begin{bmatrix} \bSig_{mm,g} & \bSig_{mo,g}\\ \bSig_{om,g} & \bSig_{oo,g}\end{bmatrix},~
        \blam= \begin{bmatrix} \blam_{m,g} \\ \blam_{o,g} \end{bmatrix}, \text{ and } \bDeltag= \begin{bmatrix} \bDelta_{m,g} \\ \bDelta_{o,g} \end{bmatrix}.
    \end{align}
    Then it is the case that:
    \begin{align}\bXo| V_g =v_g \hspace{0.15cm} \sim SN(\bmu_{o,g}, \ku\bSig_{oo,g}, \dot{\blam}_{o,g}), \nonumber    
    \end{align}
    where $\dot{\blam}_{o,g} = \frac{\bSig_{oo,g}^{-1/2}\bDelta_{o,g} }{\sqrt{ 1 - \bDelta_{o,g}^{\top}\bSig_{oo,g}^{-1}\bDelta_{o,g} } }$,
    and
    \begin{align}\bXm| \bXo=\bx^{o}, V_g =v_g \hspace{0.15cm} \sim SN(\bmu_{c,g}, \ku\bSig_{c,g}, \blam_{c,g}, \ku^{-\half}\lambda_{0,c,g}), \nonumber    
    \end{align}
    where:
    \begin{align}
     &\bmu_{c,g} = \bmu_{m,g} + \bSig_{mo,g} \bSig_{oo,g}^{-1}(\bx^{o} - \bmu_{o,g}),\hspace*{0.35cm}
     \bSig_{c,g} = \bSig_{mm,g} - \bSig_{mo,g} \bSig_{oo,g}^{-1}\bSig_{om,g},\hspace*{0.35cm} 
    \lambda_{0,c,g} = \frac{ \bDelta_{o,g}^{\top}\bSig_{oo,g}^{-1}(\bx^{(o)} - \bmu_{o,g}) }{\sqrt{1 - \bDelta_{g}^{\top}\bSigg^{-1}\bDelta_{g}} },&\nonumber\\
   & \blam_{c,g} = \frac{  \bSig_{c,g}^{-1/2} \left[\bDelta_{m,g} - \bSig_{mo,g} \bSig_{oo,g}^{-1} \bDelta_{o,g} \right] }{\sqrt{ 1 - \bDelta_{g}^{\top} \bSigg^{-1} \bDelta_{g}  } }, \text{ and } \ku = v_g + (1-v_g)\bg. &
     \end{align}
   \end{theorem}
   \begin{proof}
        The proof relies on the stochastic representation \eqref{sr}. A detailed proof can be found in \cite{sncond}. 
   \end{proof}

   \begin{theorem}
   \label{norm cond dist}
    Consider  $ \bX|g \sim CMSN(\bmug,\bSigg,\blamg,\alpha_g,\beta_g)$. Partition the random vector $\bX$ and its distribution parameters in terms of the $\bX$'s observed and missing components:
        \begin{align}
        \bX = \begin{bmatrix} \bXm\\ \bXo \end{bmatrix}, ~
        \bmug = \begin{bmatrix} \bmu_{o,g} \\ \bmu_{m,g} \end{bmatrix},~ \bOmegag = \begin{bmatrix} \bOmega_{mm,g} & \bOmega_{mo,g}\\ \bOmega_{om,g} & \bOmega_{oo,g}\end{bmatrix}, \text{ and } \bDeltag= \begin{bmatrix} \bDelta_{m,g} \\ \bDelta_{o,g} \end{bmatrix},
    \end{align} 
    where $\bOmegag = \bSigg -\bDeltag\bDeltag^{\top}$. Then it is the case that:
        \begin{align}
        \bXm | \bXo=\bx^{o}, T =t, V_g =v_g \hspace{0.15cm} \sim N(\bm{m}_{c,g} + \ku^{\half} t \bm{\gamma}_{c,g}, \ku\bOmega_{c,g}),\nonumber
        \end{align}
         where:
        \begin{align}
        &\bm{m}_{c,g} = \bmu_{m,g} + \bOmega_{mo,g}\bOmega_{oo,g}^{-1}(\bx^{o} - \bmu_{o,g}), \hspace{0.25cm}   \bm{\gamma}_{c,g}= \bDelta_{m,g} - \bOmega_{mo,g}\bOmega_{oo,g}^{-1} \bDelta_{o,g}, \hspace{0.25cm} \text{and } \hspace{0.25cm} \bOmega_{c,g} = \bOmega_{mm,g} - \bOmega_{mo,g}\bOmega_{oo,g}^{-1}\bOmega_{om,g}.&\nonumber
        \end{align}
   \end{theorem}

Theorems \ref{sn dists} and \ref{norm cond dist} assert that the distributions of $\bXm$ exist and are in closed-form when conditioned on $\bXo$, $V_g$, and/or $T$. This implies that the moments of $\bXm$ exist and can be derived similarly. This fact is crucial for carrying out the estimation procedure, given in Section \ref{estimation}.
\section{Maximum likelihood estimation}
\label{estimation}
Fitting an FMCMSN model \eqref{FMCMSN} on a random sample $\mathcal{X} = \{\bx_i\}_{i=1}^n$ is commonly achieved through the Expectation-Maximisation (EM) algorithm, a popular iterative algorithm to maximise the log-likelihood for incomplete data (\cite{em_book}). The algorithm maximises the likelihood function of a complete dataset. The complete dataset offers a log-likelihood function that produces closed-form solutions for each estimator, thereby making parameter estimation easier. The likelihood of a complete dataset involves introducing the following unobserved component membership variables $\bZ = \{ \bz_i \}_{i=1}^n$, containing vectors of binary random variables $\bz_i = [z_{i1},\dots,z_{iG}]^{\top}$ so that $z_{ig} = 1 $ if $\bx_i$ belongs to the $g^{th}$ cluster and $z_{ig} = 0 $ otherwise. That is,  the random vector $\bZ_i$ follows a multinomial distribution with parameters $\allpi = [\pi_1,\dots,\pi_G]^{\top}$. As in Section \ref{fmcmsn for missing}, $\bX_i$ is decomposed into its missing and observed vectors $\bX_i=[ \bXm_i, \bXo_i]^{\top}$. The superscripts $o$ and $m$ indicate the observed and missing parts of $\bX_i$, respectively. That is, $\bXm_i$ and $\bXo_i$ are vectors of respective lengths $p_i^m$ and $p^o_i$ with $p_i^m + p^o_i = p$. 

From the stochastic representation in \eqref{sr} and from Theorem \ref{norm cond dist}, it can be concluded that:
\begin{align*}
    &\bXm_i| \bXo_i,T_i = t_i, V_{ig} = v_{ig}, Z_{ig}=1 \sim N(\bm{m}_{c,g} + \ku^{\half} t \bm{\gamma}_{c,g}, \ku_i\bOmega_{c,g}),\\
    &\bXo_i| T_i = t_i, V_{ig} = v_{ig}, Z_{ig}=1 \sim N(\bmu_{o,g} + \ku_i^{\half}\bDelta_{o,g}, \ku_i\bOmega_{oo,g}),\\
    &\text{and} \hspace{0.25cm} T_i\sim TN(0,1).
\end{align*}
The pdf of a full observation (defined as the complete vector $\bx_i$, and latent variables $t_i, v_{ig},$ and $z_{ig}$) can be decomposed as follows:
\begin{align}
\label{miss cond pdf}
    f(\bx_i,t_i,v_i,z_{ig}) &= f(\bxm_i,\bxo_i,t_i,v_{ig},z_{ig}) &\nonumber\\
                            &= f(\bxm_i,\bxo_i|t_i,v_{ig},z_{ig})f(t_i,v_i,z_{ig}) &\nonumber\\
                            &= f(\bxm_i|\bxo_i,t_i,v_{ig},z_{ig})f(\bxo_i|t_i,v_{ig},z_{ig})f(t_i)f(v_{ig})f(z_{ig})\nonumber\\
                            &= f_N(\bxm_i;\bm{m}_{c,g} + \ku^{\half} t_i \bm{\gamma}_{c,g}, \ku_i\bOmega_{c,g})f_N(\bxo_i;\bmu_{o,g} + t_i\ku_i^{\half}\bDelta_{o,g}, \ku_i\bOmega_{oo,g})f_{TN}(t_i;0,1)f(v_{ig})f(z_{ig}),
\end{align}
 where $f_{TN}(\cdot;0, 1)$ is the pdf of a $TN(0,1)$ distribution.
Then the complete dataset with missing values is given by $\mathcal{D} = \{\bxo_i, \bxm_i, t_i, \bm{v}_i, \bz_i\}_{i=1}^n$ and its complete likelihood, using \eqref{miss cond pdf}, is given as:

\begin{align}
\label{complete likelihood}
L_c(\bm{\psi}; \mathcal{D})  = \prod_{i=1}^n \prod_{g=1}^G&\left\{ \pi_g \Big[  \alpha_g \underbrace{\bm{\phi}_p(\bx_i; \bmug + t_i\bDeltag,\bOmegag)}_{\text{good component}}f_{\text{TN}}(t_i;0,1) \Big]^{v_{ig}} \Big[  (  1  - \alpha_g)
                                                                    \underbrace{\bm{\phi}_p(\bx_i; \bmug + \beta_g^{\half}t_i\bDeltag , \beta_g\bOmegag) }_{\text{bad component}}f_{\text{TN}}(t_i;0,1) \Big]^{1-v_{ig}}\right\}^{z_{ig}},&\nonumber     \\
  = \prod_{i=1}^n \prod_{g=1}^G 
    & \left\{ 
     \pi_g 
     \Big[  
        \alpha_g ~
        \underbrace{\bm{\phi}_p(\bxm_i; \bm{m}_{c,ig} + t_i \bm{\gamma}_{c,g},\bOmega_{c,g})}_{\text{missing and good component}}~
        \underbrace{\bm{\phi}_p(\bxo_i; \bmu_{o,g} + t_i\bDelta_{o,g},\bOmega_{oo,g})}_{\text{observed and good component}}~
        f_{\text{TN}}(t_i;0,1) 
     \Big]^{v_{ig}} \right. \nonumber \\
    &\quad \left. \times 
     \Big[  
        (1 - \alpha_g) 
        \underbrace{\bm{\phi}_p(\bxm_i; \bm{m}_{c,ig} + \beta_g^{1/2}t_i\bm{\gamma}_{c,g},\beta_g\bOmega_{c,g}) }_{\text{missing and bad component}}
        \underbrace{\bm{\phi}_p(\bxo_i; \bmu_{o,g} + \beta_g^{1/2}t_i\bDelta_{o,g} , \beta_g\bOmega_{oo,g}) }_{\text{observed and bad component}}
        f_{\text{TN}}(t_i;0,1) 
     \Big]^{1-v_{ig}}
     \right\}^{z_{ig}}.
\end{align}

The corresponding complete log-likelihood function of \eqref{complete likelihood} is:

\begin{align}
    l_c(\bm{\psi}; \cD) &= l_{1}(\allpi;\cD) + l_2(\bm{\alpha};\cD) + l_3^{good}(\bmu,\bDelta,\bOmega;\cD) + l_4^{bad}(\bmu,\bDelta,\bOmega, \bm{\beta};\cD) + C,& 
\end{align}
which can be further decomposed as:
\begin{align}
\label{ll}
    l_c(\bm{\psi}; \cD) &= l_{1}(\allpi;\cD) + l_2(\bm{\alpha};\cD) \nonumber \\
    &+l_3^{m, good}(\bmu,\bDelta,\bOmega;\cD) + l_3^{o,good}(\bmu_{o},\bDelta_o,\bOmega_{oo,g};\cD) \nonumber \\
    &+l_4^{m,bad}(\bmu,\bDelta,\bOmega, \beta;\cD) + l_4^{o,bad}(\bmu_{o},\bDelta_o,\bOmega_{oo,g}, \beta;\cD) + C,& 
\end{align}
so that 
\begin{align*}
l_1(\allpi;\mathcal{D})                      = & \sum_{i=1}^n\sum_{g=1}^G z_{ig} \ln(\pi_g), \\
l_2(\bm{\alpha};\cD)                         = & \sum_{i=1}^n\sum_{g=1}^G z_{ig}( v_{ig}\ln(\alpha_g) + (1-v_{ig})\ln(1 - \alpha_g) ), \\
l_3^{m,good}(\bmu,\bDelta,\bOmega;\cD)       = &\sum_{i=1}^n\sum_{g=1}^G z_{ig}v_{ig} \ln \left [ \bm{\phi}_p(\bxm_i; \bm{m}_{c,ig} + t_i \bm{\gamma}_{c,g},\bOmega_{c,g})   \right], \\
l_3^{o,good}(\bmu,\bDelta,\bOmega;\cD)       = &\sum_{i=1}^n\sum_{g=1}^G z_{ig}v_{ig}\ln \left [ \bm{\phi}_p(\bxo_i; \bmu_{o,g} + t_i\bDelta_{o,g},\bOmega_{oo,g}) \right],\\
l_4^{m,bad}(\bmu_o,\bDelta_o,\bOmega_{oo}, \bm{\beta};\cD) = &\sum_{i=1}^n\sum_{g=1}^G z_{ig}( 1 - v_{ig} )\ln \left[\bm{\phi}_p(\bxm_i; \bm{m}_{c,ig} + \beta_g^{1/2}t_i\bm{\gamma}_{c,g},\beta_g\bOmega_{c,g}) \right],\\
l_4^{o,bad}(\bmu_o,\bDelta_o,\bOmega_{oo}, \bm{\beta};\cD) = &\sum_{i=1}^n\sum_{g=1}^G z_{ig}( 1 - v_{ig} ) \ln\left[ \bm{\phi}_p(\bxo_i; \bmu_{o,g} + \beta_g^{1/2}t_i\bDelta_{o,g} , \beta_g\bOmega_{oo,g}) \right],
\end{align*}
and $C$ is a term of constants that do not depend on any parameters. The ECM algorithm iterates between an E-step and two CM steps that alternate until there is evidence of convergence to stable parameter estimates. The steps for the $(k+1)^{th}$ iteration of the algorithm is discussed in the next section.
\subsection{ECM algorithm}
\label{ecm}
The complete log-likelihood \eqref{ll} has  the following parameters, per cluster, to estimate, namely: $\pi_g, \bmug,\bOmegag,\bDeltag,\alpha_g$ and $\beta_g$, but also five sources of unobserved values, namely: the missing components $\bxm_i$, the outlier indicators $\bm{v}_i,$ the cluster membership indicators $\bz_i$, and the truncated normal random variable $t_i$ as a consequence of employing the stochastic representation \eqref{sr} to construct the complete log-likelihood.
\subsubsection{E-step:}
\label{e_step}
 The E-step requires the expected value of \eqref{ll}, namely $Q(\bm{\psi} ) = \ev[l_c(\bm{\psi})| \cD, \bm{\psi}^{(k)}]$ using the parameter updates at the $k^{th}$ iteration, namely $\bm{\psi}^{(k)}$. The decomposition of the complete log-likelihood in \eqref{ll} implies that:
 
\begin{align}
\label{Q function}
    Q(\bm{\psi} ) & = Q_{1}(\allpi) + Q_2(\bm{\alpha}) \nonumber + Q_3^{good}(\bmu,\bDelta,\bOmega) + Q_4^{bad}(\bmu,\bDelta,\bOmega)+ C\nonumber\\   
                  & = Q_{1}(\allpi) + Q_2(\bm{\alpha}) \nonumber + Q_3^{m, good}(\bmu,\bDelta,\bOmega) + Q_3^{o,good}(\bmu_{o},\bDelta_o,\bOmega_{oo,g})\nonumber\\
                  & + Q_4^{m,bad}(\bmu,\bDelta,\bOmega, \bm{\beta}) + Q_4^{o,bad}(\bmu_{o},\bDelta_o,\bOmega_{oo,g}, \bm{\beta}) + C,
\end{align}
 where 
 
 \begin{align}
     \label{Q1}
      Q_1(\allpi)                         = & \sum_{i=1}^n\sum_{g=1}^G \zk \ln(\pi_g) \\
      \label{Q2}
      Q_2(\bm{\alpha})                    = & \sum_{i=1}^n\sum_{g=1}^G \zk \left[\vk\ln(\alpha_g) + (1-\vk)\ln(1 - \alpha_g) \right], 
 \end{align}
and the good components:
\begin{align}
\label{Q3m}
Q_3^{m,good}(\bmu,\bDelta,\bOmega)  = & -\frac{1}{2} \sum_{i=1}^n\sum_{g=1}^G \zk \vk \ln|\bOmega_{c,g}| +\frac{1}{2} \sum_{i=1}^n\sum_{g=1}^G \zk (\vktx)^{\top}\bOmega_{c,g}^{-1}\bGammacg  -\frac{1}{2} \sum_{i=1}^n\sum_{g=1}^G\zk \mathrm{tr}(\bOmega_{c,g}^{-1}\vxx)  \nonumber\\
                                      & +\frac{1}{2} \sum_{i=1}^n\sum_{g=1}^G\zk \mathrm{tr}(\bOmega_{c,g}^{-1}\vx \bm{m}_{c,ig}^{\top}) +\frac{1}{2} \sum_{i=1}^n\sum_{g=1}^G\zk \mathrm{tr}(\bOmega_{c,g}^{-1} \bm{m}_{c,ig} (\vx)^{\top})  \nonumber\\
                                      &- \frac{1}{2} \sum_{i=1}^n\sum_{g=1}^G\zk \vk \mathrm{tr}(\bOmega^{-1}_{c,g} \bm{m}_{c,ig} \bm{m}_{c,ig}^{\top}) - \frac{1}{2} \sum_{i=1}^n\sum_{g=1}^G \zk \vkt\bm{m}_{c,ig}^{\top}\bOmega_{c,g}^{-1}\bGammacg  +\frac{1}{2} \sum_{i=1}^n\sum_{g=1}^G \zk \bGammacg^{\top}\bOmega_{c,g}^{-1}\vktx\nonumber\\
                                      & - \frac{1}{2} \sum_{i=1}^n\sum_{g=1}^G \zk \vkt \bGammacg^{\top}\bOmega_{c,g}^{-1}\bm{m}_{c,ig} -\frac{1}{2} \sum_{i=1}^n\sum_{g=1}^G \zk \vktt\bGammacg^{\top}\bOmega_{c,g}^{-1}\bGammacg, \\
\label{Q3o}
Q_3^{o,good}(\bmu,\bDelta,\bOmega)  = & -\frac{1}{2} \sum_{i=1}^n\sum_{g=1}^G\zk \vk \ln|\bOmega_{oo,g}| -\frac{1}{2} \sum_{i=1}^n\sum_{g=1}^G\zk \vk d_{o,ig}^2  + \frac{1}{2} \sum_{i=1}^n\sum_{g=1}^G\zk \vkt(\bxo_i - \bmu_{o,g})^{\top}\bOmega_{oo,g}^{-1}\bDelta_{o,g} \nonumber\\
                                      & +\frac{1}{2} \sum_{i=1}^n\sum_{g=1}^G\zk \vkt \bDelta_{o,g}^{\top}\bOmega_{oo,g}^{-1}(\bx_i - \bmu_{o,g}) -\frac{1}{2} \sum_{i=1}^n\sum_{g=1}^G\zk \vktt \bDelta_{o,g}^{\top}\bOmega_{oo,g}^{-1}\bDelta_{o,g}.
\end{align}

Lastly, the expected value of the log-likelihood for the bad components are:
\begin{align}
 \label{Q4m}
Q_4^{m,bad}(\bmu,\bDelta,\bOmega, \bm{\beta}) = & -\frac{1}{2}\sum_{i=1}^n\sum_{g=1}^G p_i^m\zk ( 1 - \vk )\ln(\beta_g) -\frac{1}{2}\sum_{i=1}^n\sum_{g=1}^G \zk ( 1 - \vk )\ln|\bOmega_{c,g}|\nonumber\\
                                           & -\frac{1}{2}\sum_{i=1}^n\sum_{g=1}^G\zk\bg^{-1} \mathrm{tr}(\bOmega_{c,g}^{-1} \vcxx)
                                             +\frac{1}{2} \sum_{i=1}^n\sum_{g=1}^G\zk\bg^{-1} \mathrm{tr}(\bOmega_{c,g}^{-1}\vcxk \bm{m}_{c,ig}^{\top})\\
                                           & -\frac{1}{2}\sum_{i=1}^n\sum_{g=1}^G \zk\bg^{-1} \mathrm{tr}(\bOmega_{c,g}^{-1} \bm{m}_{c,ig} (\vcxk)^{\top})  
                                             -\frac{1}{2} \sum_{i=1}^n\sum_{g=1}^G\zk ( 1 - \vk ) \bg^{-1}\mathrm{tr}(\bOmega_{c,g}^{-1} \bm{m}_{c,ig} \bm{m}_{c,ig}^{\top})
                                           \nonumber\\
                                           & +\frac{1}{2}\sum_{i=1}^n\sum_{g=1}^G \zk(\vctxk)^{\top}\bOmega_{c,g}^{-1}\bGammacg \beta_g^{-\half} - \frac{1}{2}\sum_{i=1}^n\sum_{g=1}^G \zk ( \et - \vkt )\bm{m}_{c,ig}^{\top}\bOmega_{c,g}^{-1}\bGammacg \beta_g^{-\half}\nonumber\\
&  +\frac{1}{2}\sum_{i=1}^n\sum_{g=1}^G \zk \bGammacg^{\top}\bOmega_{c,g}^{-1}\beta_g^{-\half}\vctxk -\frac{1}{2}\sum_{i=1}^n\sum_{g=1}^G \zk ( \et - \vkt)  \bGammacg^{\top}\bOmega_{c,g}^{-1}\bm{m}_{c,ig} \beta_g^{-\half} \nonumber\\
&-\frac{1}{2}\sum_{i=1}^n\sum_{g=1}^G \zk ( \etwo - \vktt)\bGammacg^{\top}\bOmegag^{-1}\bGammacg~, \\
\label{Q4o}
Q_4^{o,bad}(\bmu,\bDelta,\bOmega, \bm{\beta}) = & -\frac{1}{2}\sum_{i=1}^n\sum_{g=1}^G \zk ( 1 - \vk )p_i^o\ln(\beta_g) -\frac{1}{2}\sum_{i=1}^n\sum_{g=1}^G \zk ( 1 - \vk )\ln|\bOmega_{oo,g}| \nonumber\\
                                           & -\frac{1}{2}\sum_{i=1}^n\sum_{g=1}^G \zk ( 1 - \vk ) d_{o,ig}^2\bg^{-1} +\frac{1}{2} \sum_{i=1}^n\sum_{g=1}^G \zk ( \et - \vkt ) (\bx_i - \bmu_{o,g})^{\top}\bOmega_{oo,g}^{-1}\bDelta_{o,g} \beta_g^{-\half}\nonumber\\
                                           & +\frac{1}{2}\sum_{i=1}^n\sum_{g=1}^G \zk ( \et - \vkt ) \bDelta_{o,g}^{\top}\bOmega_{oo,g}^{-1}(\bx_i - \bmu_{o,g})\beta_g^{-\half} -\frac{1}{2}\sum_{i=1}^n\sum_{g=1}^G \zk ( \etwo - \vktt ) \bDelta_{o,g}^{\top}\bOmega_{oo,g}^{-1}\bDelta_{o,g}. 
\end{align}

The following expectations found in the function $Q$ given in \eqref{Q function} are important as they are used to cluster an observation according and identify it as a good or bad point, respectively (further details are discussed in Section \ref{cluster_and_outlier}): 
\begin{align*}
    \zk  &= \mathbb{E}\left[Z_{i,g}  |  \bxo_i ,\bm{\psi}^{(k)}  \right] = \frac{\pi_g^{(k)} f_{CMSN} \left(\bxo_i;\bmu^{(k)}_{o,g}, \bSig^{(k)}_{oo,g},\dot{\blam}^{(k)}_{o,g},\bg^{(k)}\right)  }{ f_{FMCMSN}(\bxo_i;\bm{\psi}^{(k)}_{o}) },
\end{align*}
and
\begin{align*}
    \vk  &=  \mathbb{E}\left[V_{i} | Z_{i,g}, \bx^{o}_i,  \bm{\psi}^{(k)}\right] = \frac{\alpha_g^{(k)} f_{MSN} \left(\bxo_i;\bmu_{o,g}^{(k)}, \bSig_{oo,g}^{(k)},\dot{\blam}^{(k)}_{o,g}\right)  }{f_{CMSN} \left(\bxo_i;\bmu_{o,g}^{(k)}, \bSig_{oo,g}^{(k)},\dot{\blam}^{(k)}_{o,g}, \bg^{(k)}\right)  },
\end{align*}
with $\dot{\blam}_{o,g}^{(k)} =  \frac{\bSig_{oo,g}^{(k) -\half}\bDelta_{o,g}^{(k)} }{\sqrt{ 1 - \bDelta_{o,g}^{^{(k)} \top}\bSig_{oo,g}^{^{(k)} -1}\bDelta_{o,g}^{(k)} } }$ and $\bm{\psi}^{(k)}_{o} =\left \{  \bmu^{(k)}_{o,g}, \bSig^{(k)}_{oo,g},\dot{\blam}^{(k)}_{o,g},\bg^{(k)}    \right \}_{g=1}^{G}$. The expectations $ \vxx, \vx, \vktx$, $\vcxx$, $\vcxk$, $\vctxk$, and $\vkt, \et,$ and $\etwo$ are in closed-form as equations \eqref{t_given_v1} -- \eqref{vctx} that can be found in the Appendix.


\subsubsection{CM-step 1:}
The first CM step calculates the updates $\bm{\psi}^{(k+1)}$ by maximising $Q$ in \eqref{Q function}. Notice that $Q_1$ and $Q_2$ can be maximised independently, leading to the updates:
\begin{align*}
\pi^{(k+1)}_g         = & \frac{ n^{(k)}_g }{n},\\
\alpha^{(k+1)}_g      = & \frac{\displaystyle\sum_{i=1}^n\zk\vk }{n^{(k)}_g},
\end{align*}
where $n^{(k)}_g = \displaystyle\sum_{i=1}^{n} \zk$ is the effective size of the $g^{th}$ cluster. Each observation may have incomplete components. During the E-step, the expected values of the missing components are computed and substituted into the corresponding positions of the vector where the missingness occurs. For notational convenience, we denote the resulting complete vector as:
\begin{align*}
    \vhk &= \ev[V_i\bX_i|\bxo_i, \bm{\psi}^{(k)}] = [\vk\bxo_i, \vx]^{\top} \\
    \uk &= \ev[V_iT_i\bX_i|\bxo_i, \bm{\psi}^{(k)}] = [\vkt\bxo_i, \vktx]^{\top} \\
    \vhck &= \ev[(1-V_i)\bX_i|\bxo_i, \bm{\psi}^{(k)}] = [(1-\vk)\bxo_i, \vcxk]^{\top} \\
    \uck &= \ev[(1-V_i)T_i\bX_i|\bxo_i, \bm{\psi}^{(k)}] = [(\et-\vkt)\bxo_i, \vctxk]^{\top} \\
    \vHk &= \ev[V_i\bX_i\bX_i^{\top}|\bxo_i, \bm{\psi}^{(k)}] = \begin{bmatrix} \vk\bxo_i (\bxo_i)^{\top} &    \bxo_i (\vx)^{\top} \\ \vx (\bxo_i)^{\top} & \vxx  \end{bmatrix}, \text{ and}\\
    \vHck &= \ev[(1 - V_i)\bX_i\bX_i^{\top}|\bxo_i, \bm{\psi}^{(k)}] = \begin{bmatrix} (1-\vk)\bxo_i (\bxo_i)^{\top} &    \bxo_i (\vcxk)^{\top} \\ \vcxk (\bxo_i)^{\top} & \vcxx  \end{bmatrix}.
\end{align*} It is important to note that this notation does not imply that the observed and expected components are necessarily divided evenly within the vector.
Functions $Q_3$ and $Q_4$ are maximised with respect to the $g^{th}$ location vector $\bmug$ and $\bDeltag$: 

\begin{align*}
\bmu^{(k+1)}_g    = & \frac{ C^{(k)} \left[ \displaystyle\sum_{i=1}^n \zk( \vhk + \frac{1}{\bg^{(k)} } \vhck ) \right] - A^{(k)}\left[ \displaystyle\sum_{i=1}^n \zk (\uk + \frac{ 1 }{\bg^{(k)  \half} }\uck ) \right]}{ B^{(k)}C^{(k)} - (A^{(k)})^2 },\\
\bDelta^{(k+1)}_g = & \frac{ B^{(k)} \left[ \displaystyle\sum_{i=1}^n \zk (\uk + \frac{ 1 }{\bg^{ (k) \half}}\uck ) \right] - A^{(k)} \left[\displaystyle\sum_{i=1}^n \zk( \vhk + \frac{1}{\bg^{(k)} } \vhck )  \right] } { B^{(k)}C^{(k)} - (A^{(k)})^2   },
\end{align*}
where $A^{(k)} = \displaystyle\sum_{i=1}^n \zk \left(\vkt + \frac{ \et - \vkt }{\bg^{ (k)  \half}} \right)$, $B^{(k)} =\displaystyle\sum_{i=1}^n\zk\left( \vk + \frac{ 1 - \vk }{\bg^{(k)} } \right)$, and $C^{(k)} = \displaystyle\sum_{i=1}^n \zk \etwo$.
With $\bmug$ and $\bDeltag$ updated, they are used to update $\bOmegag$ as follows:
\begin{align*}
\bOmega_g^{(k+1)}     = &\frac{1}{n_g^{(k)} } \Big\{
                                \sum_{i=1}^{n} \zk  \left(\vHk + \frac{1}{\beta_g^{(k)} } \vHck \right)
                                -\sum_{i=1}^{n} \zk \bmug^{(k+1)}\left(\vhk+ \frac{1}{\beta_g^{(k)} } \vhck \right)^{\top}
                                -\sum_{i=1}^{n} \zk \left(\vhk+ \frac{1}{\beta_g^{(k)} } \vhck \right)\bmug^{(k+1)\top}&\nonumber\\
                                &+B^{(k)}\bmug^{(k+1)}\bmug^{(k+1)\top} - \sum_{i=1}^{n} \zk\bDeltag^{(k+1)}\left(\uk + \frac{1}{\beta_g^{ (k) \half }}\uck \right)^{\top} 
                                 - \sum_{i=1}^{n} \left(\uk + \frac{1}{\beta_g^{ (k) \half}}\uck \right)\bDeltag^{(k+1)\top}\nonumber&\\
                                &+                A^{(k)} \bDeltag^{(k+1)}\bmug^{(k+1)\top}
                                 +                A^{(k)}\bmug^{(k+1)}\bDeltag^{(k+1)\top}
                                 +                C^{(k)}\bDeltag^{(k+1)}\bDeltag^{(k+1)\top}
                                   \Big\}&.
\end{align*}


\subsubsection{CM-step 2:}

A closed-form estimator for $\beta_g$ involves solving a quadratic equation with two distinct solutions. Since $\beta_g>1$ the positive solution is chosen and the negative solution is ignored. Unfortunately, it cannot be guaranteed that the positive solution will be larger than one. Thus, $\beta_g$ is updated as a choice between the positive solution and a pre-determined lower threshold value $\beta^*>1$ to ensure $\beta^{(k+1)}_g >1$:
\begin{align*}
\beta^{(k+1)}_g       = & \max\left\{\beta^*,  \left( 
                                          \frac{D^{(k)}}{2} + \sqrt{ \left( \frac{D^{(k)}}{2} \right)^2  
                                          +\frac{ \displaystyle\sum_{i=1}^n\zk d_{ig} ^{(k)}   }{\displaystyle\sum_{i=1}^n\zk \vk } } \right)^2 \right\}, 
\end{align*}
where
\begin{align*}
    d_{ig} ^{(k)} &= \mathrm{tr}  \Big\{ (\bOmegag^{(k+1)})^{-1}  \big[ \vHck
                                 - \bmu^{(k+1)}_g(\vhck )^{\top}    
                                 - (\vhck )\bmu^{(k+1)\top}_g  
                                 + (1 - \vk)\bmu^{(k+1)}_g\bmu_g^{(k+1)\top} 
                                 - \bDelta^{(k+1)}_g( \uck )^{\top} \nonumber\\
                               & -\uck \bDelta^{(k+1)\top}_g\quad + (1 - \vkt)\bDelta_g^{(k+1)}\bmu_g^{(k+1)\top}
                                 + (1 - \vkt)\bmu^{(k+1)}_g\bDelta_g^{(k+1)\top}
                                 + (\etwo - \vktt)\bDelta^{(k+1)}_g\bDelta^{(k+1)\top}_g \big]
                                \Big\},
\end{align*}
 and
 \begin{align*}
 D^{(k)} = \frac{\displaystyle\sum_{i=1}^n\zk\bDeltag^{(k+1)\top}(\bOmegag^{(k+1)})^{-1}(\bmug^{(k+1)} -\uck)  }{p\displaystyle\sum_{i=1}^n\zk\vk } .
 \end{align*}
A value of $\beta^* =1.001$ has shown to be a suitable lower threshold value (\cite{tong2024missing}).
\newline
\newline

\subsection{Initialisation and convergence}
The ECM algorithm relies on sufficient starting points for the best possible chance for the algorithm to converge to the log-likelihood's global maximum and avoid one of possibly several local maxima (\cite{local_max, local_max2}). Popular initialisation techniques rely on variants on the EM algorithm, such as the CEM, SEM, and small-EM algorithm techniques, which in turn rely on results from other clustering techniques, such as hierarchical clustering, $k$-means, and $k$-medoids or random partitions (\cite{em_starts}). Furthermore, the dataset contains outliers and is skewed which adds more complexity to motivating for a suitable starting point. From surveyed literature, a stable point of initialisation with asymmetric data includes the method of moments using clustering results from either $k$-means and $k$-medoids. Each of the inflation parameters $\{ \beta_g \}^G_{g=1}$ is set as a value close to 1 as a conservative approach until the algorithm guides it away from 1.

The log-likelihood of the observed dataset increases monotonically at each iteration of the ECM algorithm. However, the algorithm may run into a local maximum before it finds a global maximum, in which case the initial stability is temporary -i.e. the observed log-likelihood values are stable only for a few iterations before they increase again. The Aitken acceleration criterion is thus used in this paper to determine whether the algorithm has converged to its asymptotic value. Let $l_o^{(k)}$ denote the observed log-likelihood at the $k^{th}$ iteration. Then the Aitken acceleration criterion is given as:
\begin{align}
    a^{(k+1)} = \frac{l_o^{(k+2)} - l_o^{(k+1)}}{l_o^{(k+1)} - l_o^{(k)}}.
\end{align}
Then the estimated asymptotic observed log-likelihood at the $k^{th}$ iteration, say $(l_o^{\infty})^{(k)}$ is:
\begin{align}
    (l_o^{\infty})^{(k)} = l_o^{(k+1)} + \frac{ l_o^{(k+2)} - l_o^{(k+1)}  }{1 - a^{(k+1)}}.
\end{align}
The EM algorithm is therefore considered to have converged if $(l_o^{\infty})^{(k)} - l_o^{(k+1)} < \epsilon$ where $\epsilon>0$ is a small number (\cite{convergence_aitken}).

\subsection{Automatic clustering and outlier detection}
\label{cluster_and_outlier}
The construction of the complete-data log-likelihood \eqref{ll} allows one to determine cluster membership through their respective maximum aposteriori (MAP) probabilities. That is, setting $\hat{z}^{(f)}_{ig}$ to be the value of $\zk$ at convergence of the ECM algorithm, an observation $\bxo_i$ is deemed part of the $g^{th}$ cluster if $\hat{z}_{ig} = \max\{\hat{z}^{(f)}_{ij} \}_{j=1}^G$. The second indicator, $v_{i}$  distinguishes between ‘good’ and ‘bad’ points in each cluster. This model offers outlier detection capability via the following aposteriori probability for an observation $\bx$ using the estimated parameters:
\begin{align}
    \hat{v}_{ig} = \frac{ \hag_g f_{MSN}(\bx_i;\hmg_g,\hsg_g,\hlg_g)  }{f_{CMSN}(\bx_i;\hmg_g,\hsg_g,\hlg_g,\hag_g, \hbg_g)},
\end{align}
where $\hag_g, \hbg, \hmg_g,\hsg_g$ and $\hlg_g$ are $\alpha^{(k)}_g, \beta^{(k)}_g, \bmu^{(k)}_g,\bSigg^{(k)}$ and $\blamg^{(k)}$ at convergence. For $\hat{v}_{ig}<0.5,~ \bx_i$ is considered an outlier. Conveniently, this aposteriori probability is automatically calculated as part of the ECM algorithm's $k^{th}$ iteration. Thus, setting $\widehat{v}_{ig}$ to be the last value of $\vk$ at convergence of the ECM algorithm provides an output of whether a point can be classified as an outlier or not (\cite{convergence_aitken}).

\section{Simulation experiments}
\label{simulation}
The performance of the FMCMSN model is assessed via its ability to correctly cluster an observation into the component of origin, and its ability to detect outliers in the dataset while simultaneously imputing missing values through the proposed algorithm. It is assessed under the effect of sample size, proportion of outliers present, and percentage of missing rows in the sample. 

\subsection{Competitors}
Within the model-based clustering framework, the components of the finite mixtures of the contaminated multivariate normal (FMCMN) are the symmetric special case of the FMCMSN and as such also has the capability to automatically detect outliers and has been extended to handle missing values at random (\cite{tong2024missing}). The multivariate Student t (Mt) and multivariate skew Student t (MSt) also address heavier tailed components, with the latter also accounting for skewness. The \textbf{MixtureMissing} package in \textsf{R} contains functions for fitting a finite mixture of the Mt (FMMt) and FMCMN models for incomplete data. Algorithms for fitting mixtures of the MSN (FMMSN), and MSt (FMMSt) distributions for incomplete data have been developed and thus are also competitors for cluster performance when clusters are skewed (\cite{censmfm, archive}).

As for outlier detection, only FMCMN and FMMt models are viable competitors to compare against the FMSCN model as they are the only models with some kind of outlier detection capabilities. The FMCMN model is a natural competitor, as it is the symmetric case of the FMCMSN model. The FMMt distribution can identify outliers through the Mahalanobis distance. Specifically, the Mahalanobis squared distance: 
\begin{align*}
    d^2(\bx;\bmu,\bSig) = (\bx - \bmu)^{\top} \bSig^{-1}(\bx - \bmu)
\end{align*}
follows a chi-squared distribution with $p$ degrees of freedom ($\chi^2_p$) when $\bx$ is an observation from a $N(\bmu,\bSig)$ distribution (\cite{ghorbani2019mahalanobis}). $D^2(\bx;\bmu,\bSig)$ can also be used when the data is generated by a FMMt model (\cite{mitchell1985mahalanobis}). The statistic:
\begin{align*}
    D_i^2 = \sum_{g=1}^G \hat{z}_{ig}d^2(\bx_i;\hmg_g,\hsg_g)
\end{align*}
flags $\bx_i$ as an outlier when it lies beyond a specified percentile $\chi^2_p$ percentile. The choice of the percentile is subjective, and from the literature surveyed, the 95th percentile is often suggested as a suitable threshold (\cite{mahalanobis_t}).

\subsection{Simulation experiment design}
The design outline and execution is similar to that by \citeyearbracket{tong2024missing}. The experiment consists of two parts, namely: part A and part B. Part A considers data from a two component mixture model, with outliers generated. The clustering performance and outlier detection capabilities of the proposed model and its competitors are compared while controlling for possible confounding factors. Part B illustrates two distinct outlier scenarios in one simulation study. Specifically, we introduce two types of outliers selected at random: the first aligned with the direction of skewness, and the second positioned near the bulk of the data but deliberately not in the direction of maximal directional skewness (\cite{chem2008}). This design aims to evaluate the outlier capabilities of the proposed model against its competitors in a high dimensional setting.

\subsubsection{Simulation experiment: Part A}
The first part of the experiment constructs a bivariate sample generated under the following cases for 2 clusters ($G=2$):
\begin{itemize}
    \item[(a)] FMMSt with $\nu_1$ = 4 and $\nu_2$ = 10 degrees of freedom.
    \item[(b)] FMCMSN with $\alpha_1$ = 0.9, $\alpha_2$ = 0.8, $\beta_1$ = 20, and $\beta_2$ = 30.
    \item[(c)] FMMSN with 1\% of points are randomly replaced by (0, $x^{*}_{i2}$), where $x^{*}_{i2}$ is a simulated point from a continuous uniform distribution on the interval (10, 15).
    \item[(d)] FMMSN with 5\% of points randomly replaced by noise from a continuous uniform distribution on the square $(0,10)^2$.
    \item[(e)] FMMSN with 20\% of points randomly replaced by noise from a continuous uniform distribution on the square $(0,10)^2$.
\end{itemize}
For each case the following parameters are fixed:
\begin{align*}
    \pi_1 = 0.3 , \hspace{0.2cm}
    \pi_2 = 0.7,\hspace{0.2cm} 
    \bmu_2 = \begin{bmatrix} 0\\3 \end{bmatrix},\hspace{0.2cm}
    \bSig_1 = \begin{bmatrix} 2 & -1 \\-1 & 2 \end{bmatrix},\hspace{0.2cm}
    \bSig_2 = \begin{bmatrix} 2 & 1 \\1 & 2 \end{bmatrix}, \hspace{0.2cm}
    \blam_1 = \begin{bmatrix} 3\\5 \end{bmatrix},\hspace{0.2cm}
    \blam_2 = \begin{bmatrix} 4\\2 \end{bmatrix}.
\end{align*}

Clustering performance of the FMCMSN model is assessed while accounting for cluster proximity and sample size. Data generation in each case is simulated for a close and far proximity, achieved by setting $\bmu_1 = [0 \hspace{0.1cm}, -1] ^{\top}$  and $\bmu_1 = [0 \hspace{0.1cm}, -3] ^{\top}$, respectively. For each proximity level, two sample sizes are considered: a small sample ($n = 300$) and a large sample ($n = 800$). Within each proximity - sample size combination, missing values are then introduced at random using the \textbf{mice} package in \textsf{R}. The proportion of rows containing missing values is varied across 0\%, 20\%, 40\%, 60\%, and 80\%. For every combination of proximity, sample size, and missingness proportion, clustering performance is quantified using the Adjusted Rand Index (ARI). Accuracy rates, true positive rates (TPRs) and false positive rates (FPRs) are also reported to assess the models' outlier detection capabilities. Here, an accuracy rate is defined as the proportion of points in the data correctly classified by the model, a TPR is defined as the proportion of outliers correctly detected by the model, and a FPR is the number of good points incorrectly identified as outliers over the total number of good points.

\begin{figure}[H]
    \centering
    \includegraphics[trim={0cm 0cm 0cm 1cm}, clip, width=.9\textwidth]{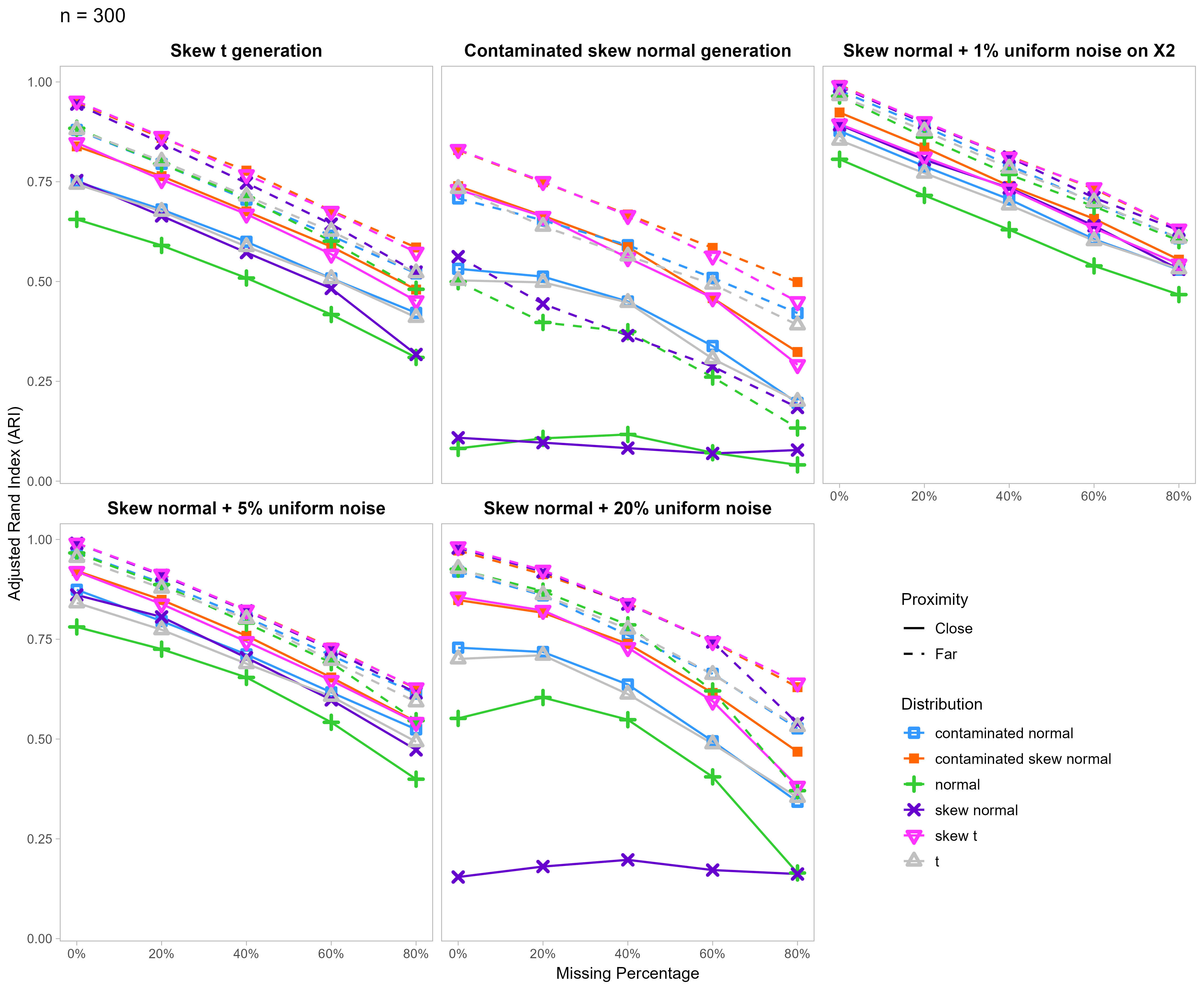}
    \caption{Average ARI values for $n$=300 obtained for each model under the four data-generating processes (a)–(d), across two levels of cluster overlap and for varying percentages of incomplete rows in the dataset.}
    \label{ari_300}
\end{figure}

\begin{figure}[H]
    \centering
    \includegraphics[trim={0cm 0cm 0cm 1cm}, clip, width=.9\textwidth]{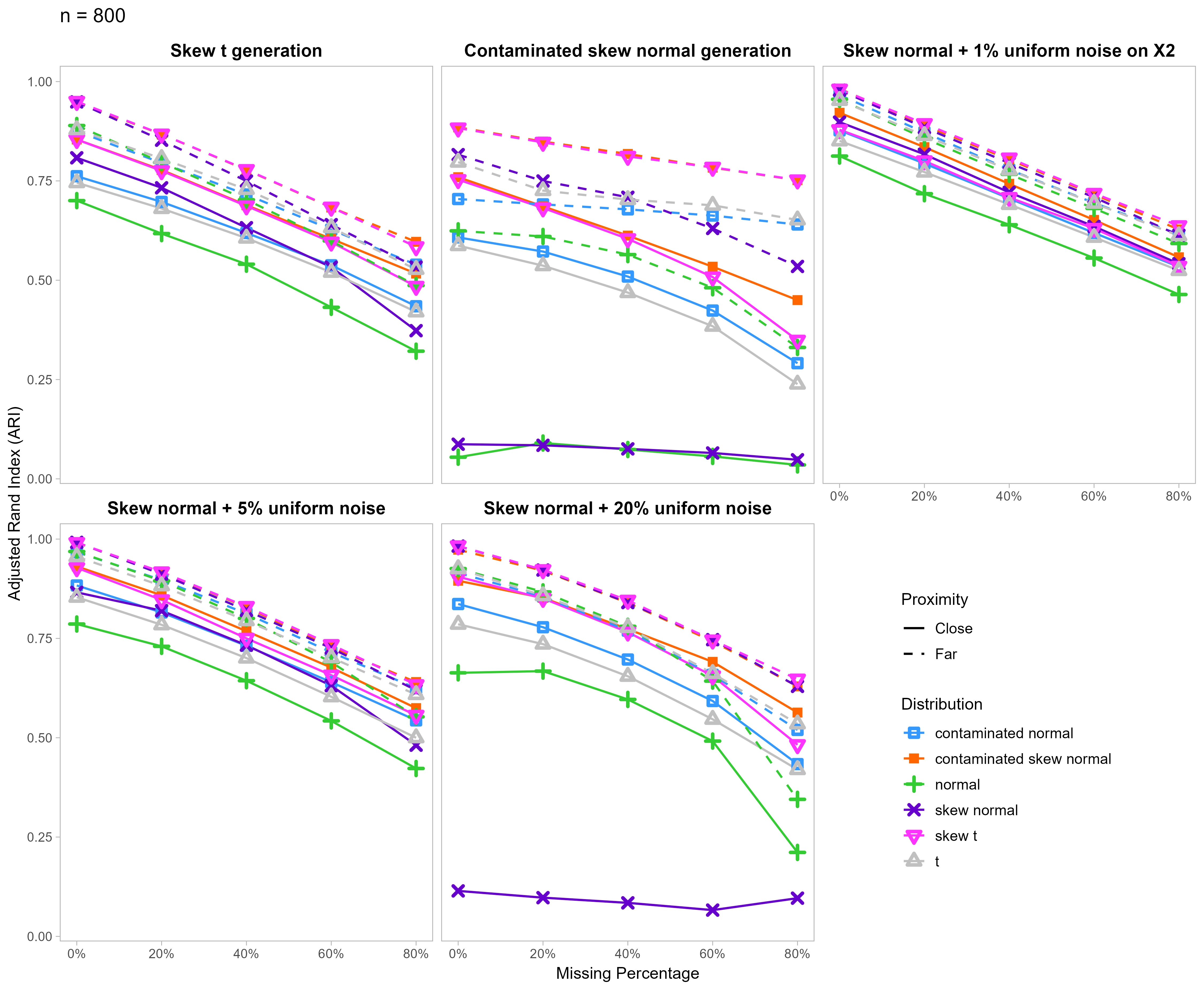}
    \caption{Average ARI values for $n$=800 obtained for each model under the four data-generating processes (a)–(d), across two levels of cluster overlap for varying percentages of incomplete rows in the dataset.}
    \label{ari_800}
\end{figure}

\begin{figure}[H]
    \centering
    \includegraphics[trim={0cm 0cm 0cm 1cm}, clip, width=\textwidth]{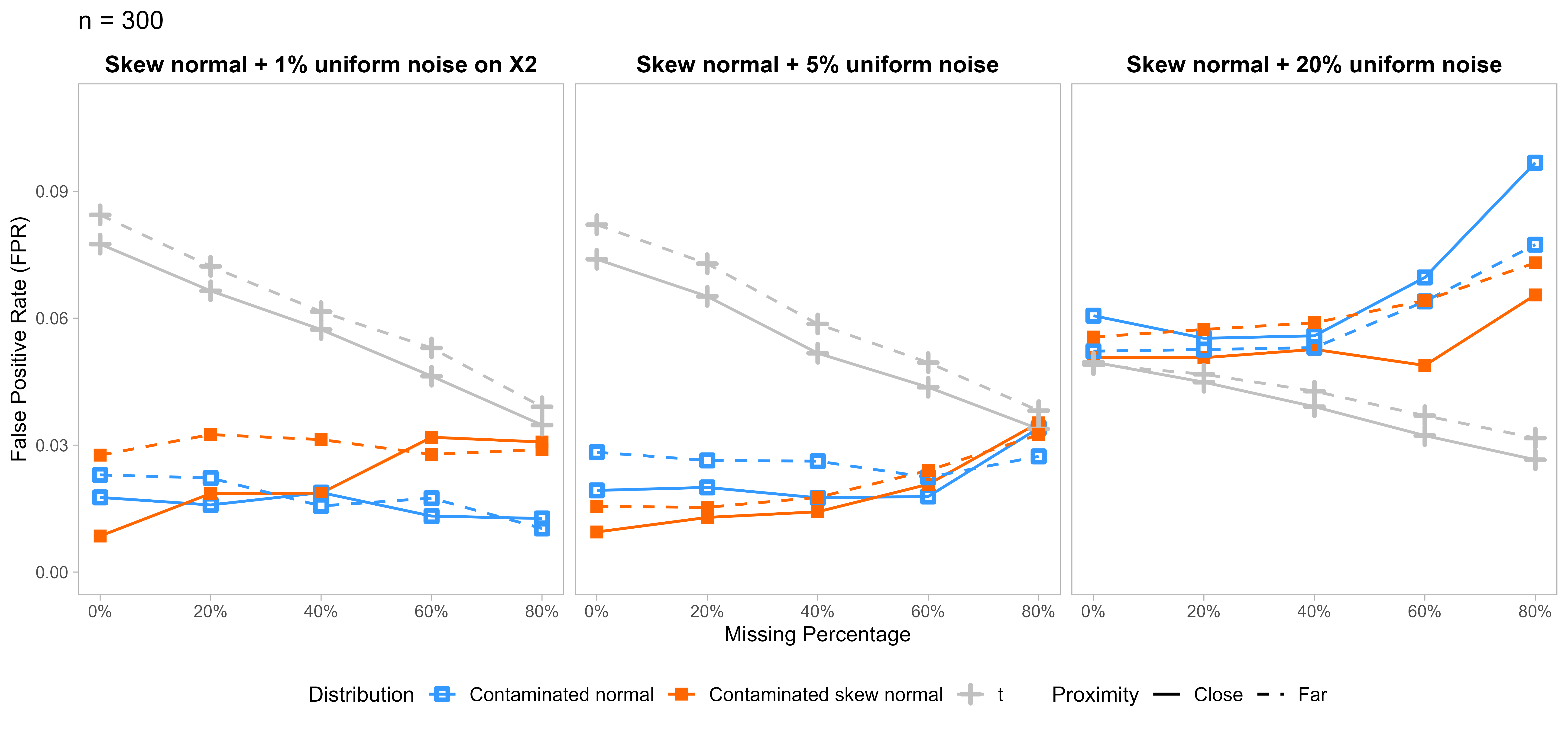}
    \caption{Average FPR values for $n=300$. Results are shown for the FMMt, FMCMN, and FMCMSN models across varying percentages of incomplete rows and across two levels of cluster proximity.}
    \label{fpr_300}
\end{figure}

\begin{figure}[H]
    \centering
    \includegraphics[trim={0cm 0cm 0cm 1cm}, clip, width=\textwidth]{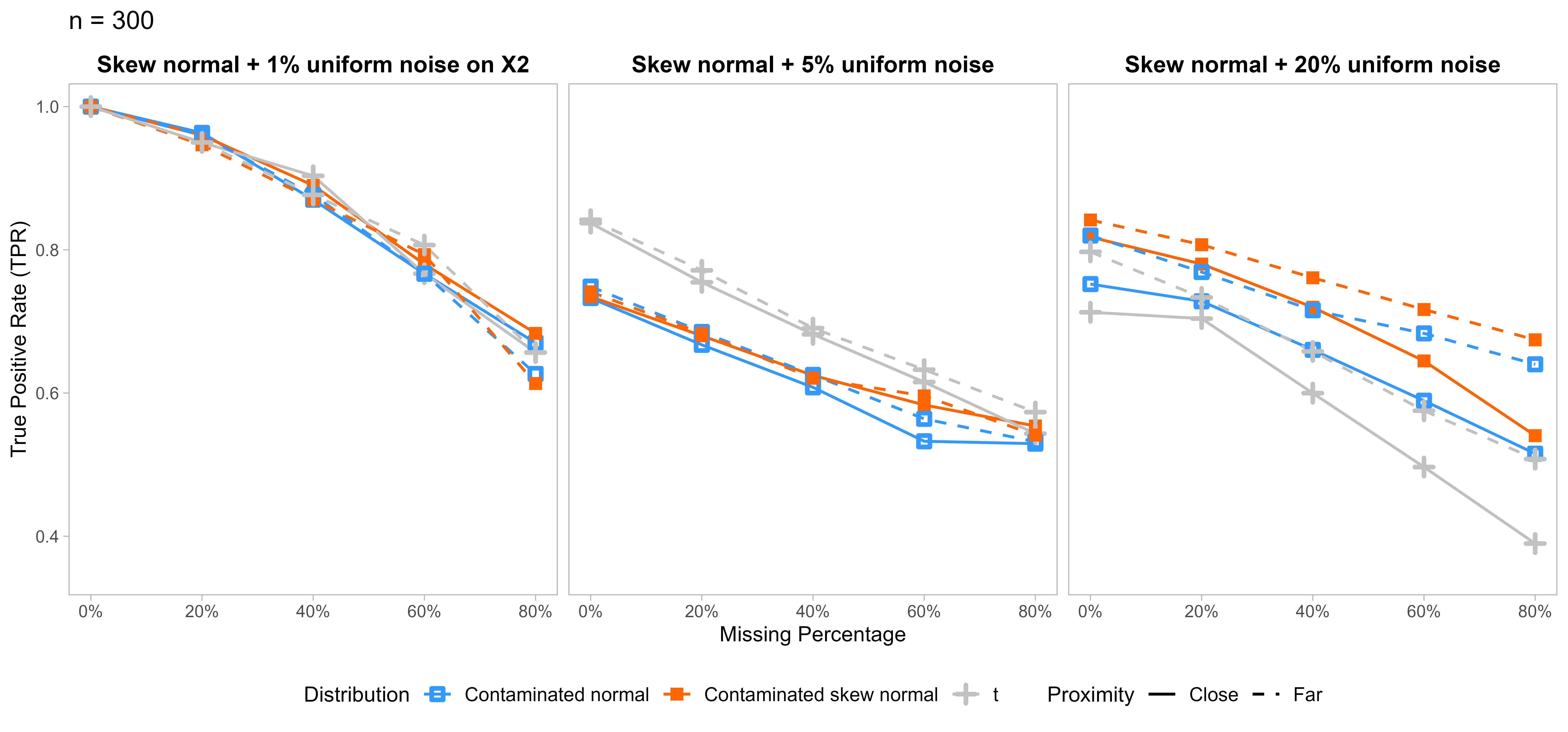}
    \caption{Average TPR values for $n=300$. Results are shown for the FMMt, FMCMN, and FMCMSN models across varying percentages of incomplete rows and across two levels of cluster proximity.}
    \label{tpr_300}
\end{figure}

\begin{figure}[H]
    \centering
    \includegraphics[trim={0cm 0cm 0cm 1cm}, clip, width=\textwidth]{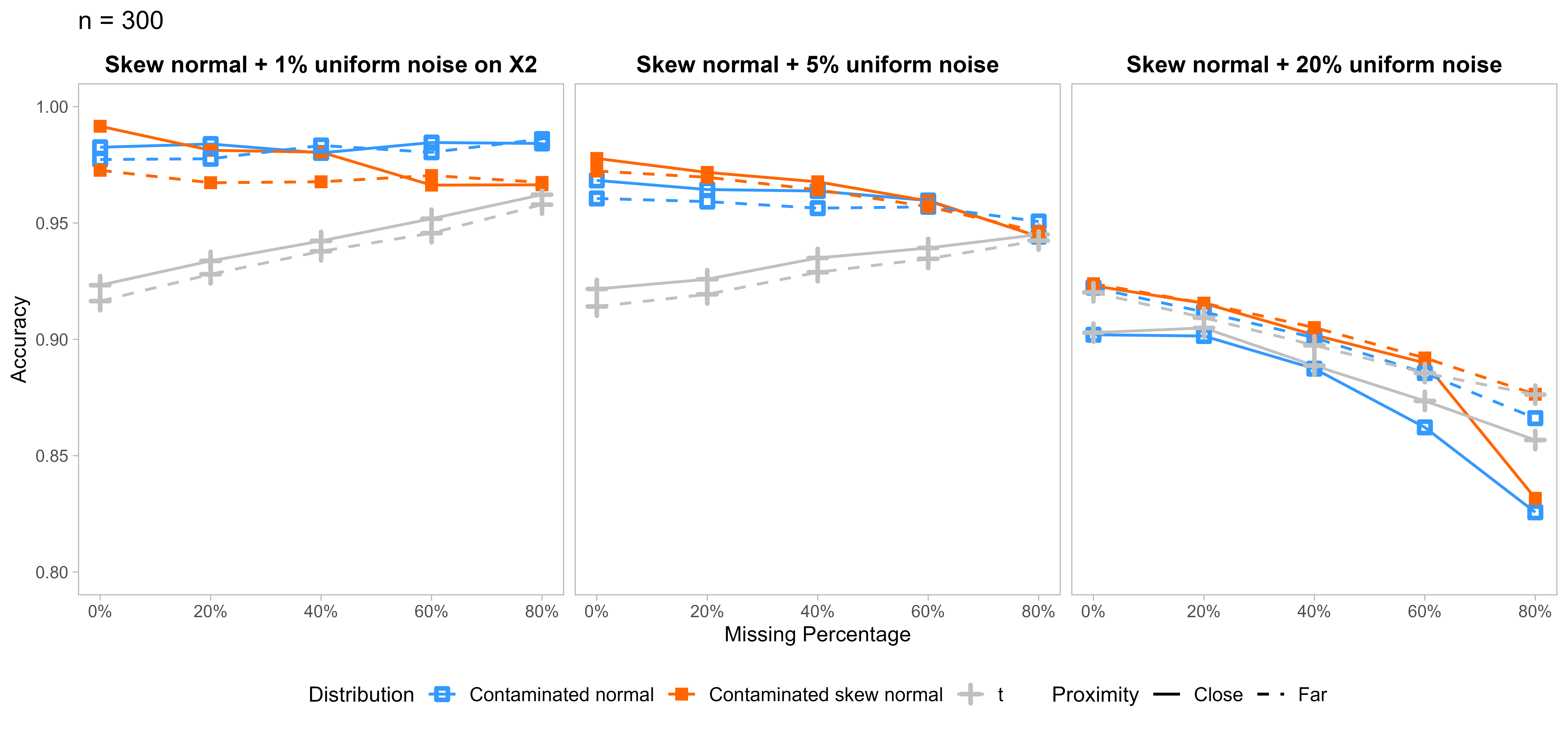}
    \caption{Proportion of accurately classified points for $n=300$. Results are shown for the FMMt, FMCMN, and FMCMSN models across varying percentages of incomplete rows and across two levels of cluster proximity.}
    \label{acc_300}
\end{figure}

\begin{figure}[H]
    \centering
    \includegraphics[trim={0cm 0cm 0cm 1cm}, clip, width=\textwidth]{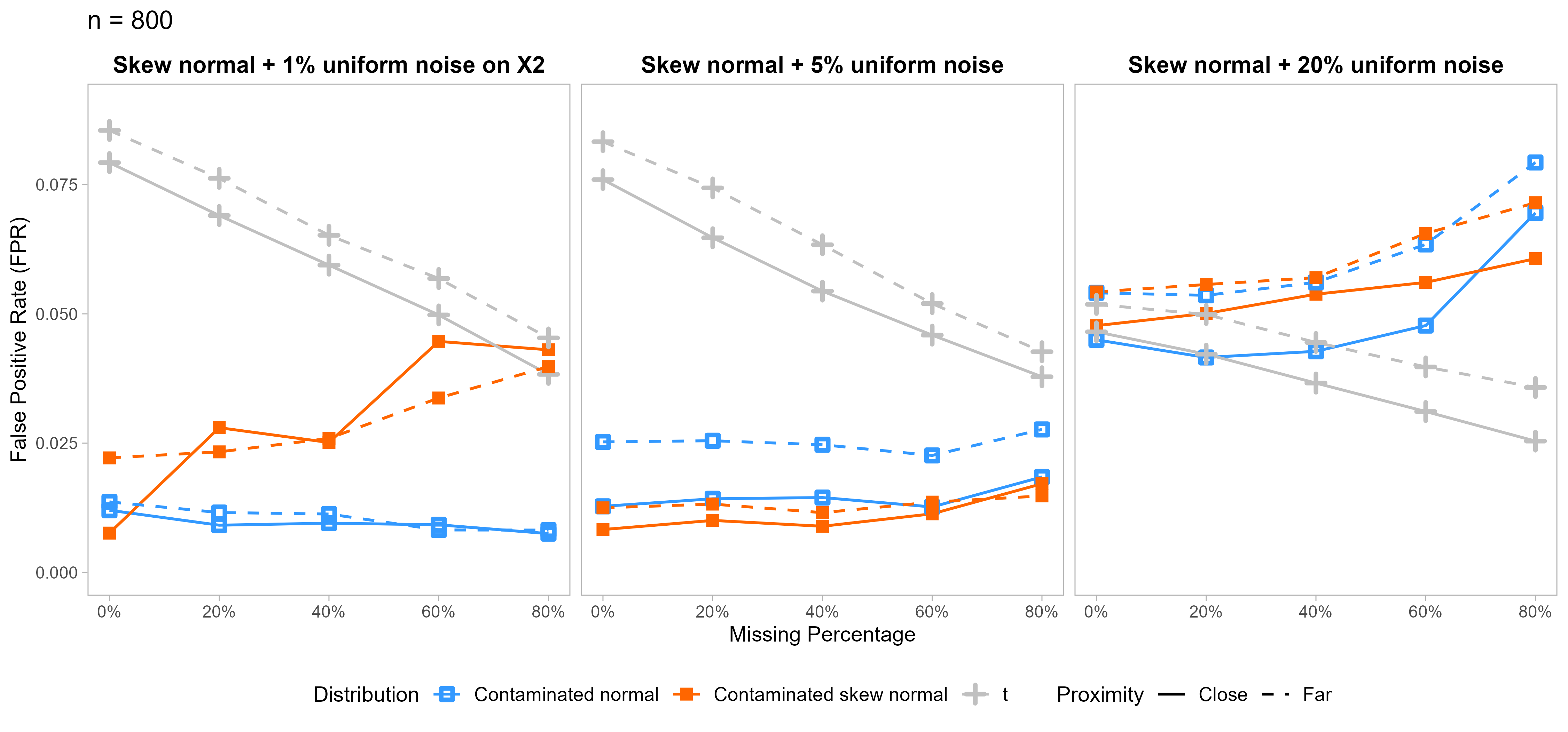}
    \caption{Average FPR values for $n=800$. Results are shown for the FMMt, FMCMN, and FMCMSN models across varying percentages of incomplete rows and across two levels of cluster proximity.}
    \label{fpr_800}
\end{figure}

\begin{figure}[H]
    \centering
    \includegraphics[trim={0cm 0cm 0cm 1cm}, clip, width=\textwidth]{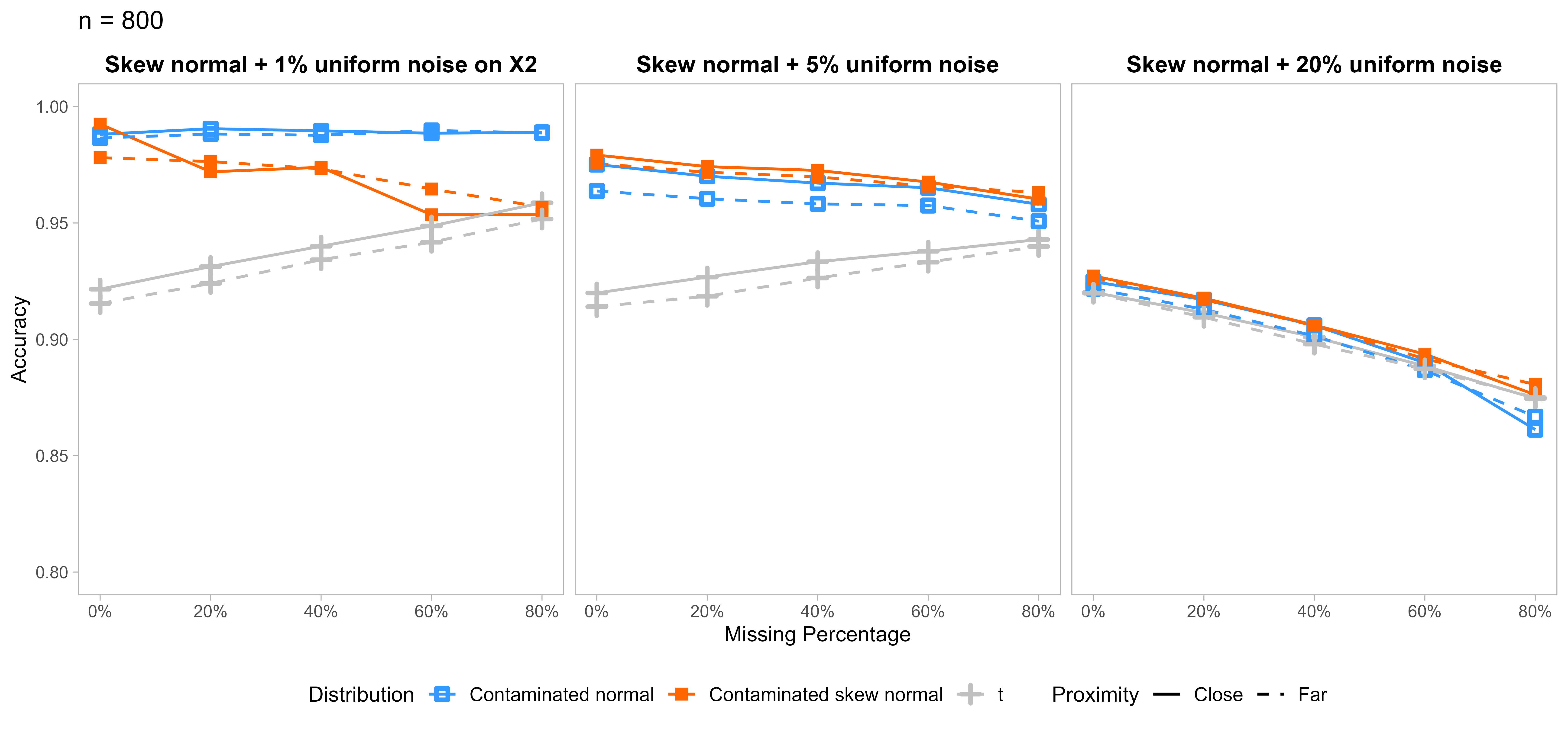}
    \caption{Proportion of correctly-classified points for $n=800$. Results are shown for the FMMt, FMCMN, and FMCMSN models across varying percentages of incomplete rows and across two levels of cluster proximity.}
    \label{acc_800}
\end{figure}

For each scenario considered, 100 samples are generated from which an average ARI, TPR, and FPR are computed (see Figures \ref{ari_300}--\ref{fpr_800}). The average ARI values suggest that the FMCMSN performs the best overall, with the FMMSt distribution a close second. When cluster proximity is set to be large, the average ARIs of FMMSt and FMCMSN models overlap perfectly. The FMMSt is expected to be a close competitor to the FMCMSN model (\cite{tong2024missing}), and the effect of large proximity and large sample sizes further aid the FMMSt model's performance against the FMCMSN model. The differences in performance are better highlighted when there is a more prominent cluster overlap. The average ARIs show that the FMMSt trend falls off more quickly than the FMCMSN trend. In other words, as the percentage of incomplete rows grows, the FMMSt stops approximating the underlying distribution as effectively as the FMCMSN. These findings agree with the expectations of the models under comparison. In agreement with the findings of \citeyearbracket{tong2024missing}, the models that perform better after the FMCMSN and FMMSt models are the FMCMN and FMMt models. That is, the cluster-wise symmetric special cases of the FMCMSN and FMMSt models are the next best contenders, though their performance is outshone by their skewed counterparts. Last in the ranking is the FMMN and FMMSN models. Particularly poor performance is noted when the data generating process is a FMCMSN model, and FMMSN with 20\% noise. That is, it is a real disadvantage to the FMMN and FMMSN models that do not have the capabilities to account for outliers or anomalous values. Overall, there is a clear downward trend in clustering performance as the proportion of missing rows in the dataset increases as expected. This trend acts independently of sample size and cluster proximity.\newline
\newline
With regard to outlier detection, Figures \ref{fpr_300} and \ref{fpr_800} reflect that under a few points of contamination, the ability of all the models to correctly identify bad points decreases as the missing proportion increases. This can be explained by recognising that a bad point that has missing values is more difficult to identify as a good point. Further, missing values also remove the number of skewed points, which in turn, underestimates the skewness parameters, making it more likely for the point to be classified as a bad point rather than as a good point belonging to the cluster. Overall, performance shows benefit at balancing both clustering and outlier detection.  The trends in the FPR and TPR for the FMCMN model mimics that of the FMCMSN model, which is to be expected. At first glance, it appears that the FMMt's TPR outperforms the FMCMN and FMCMSN models at outlier identification when 5\% of noise is present in the data, but the FPR in Figures \ref{fpr_300} and \ref{fpr_800} suggest that the model is instead biased towards classifying a point as an outlier which impedes the accuracy of the model, as is evident in Figures \ref{acc_300} and \ref{acc_800}. Consequently, the FMCMSN model outperforms the other two overall with the highest accuracy rates for increasing percentage of outliers in the data. 

\begin{figure}[H]
    \centering
    \includegraphics[trim={0cm 0cm 0cm 1cm}, clip, width=\textwidth]{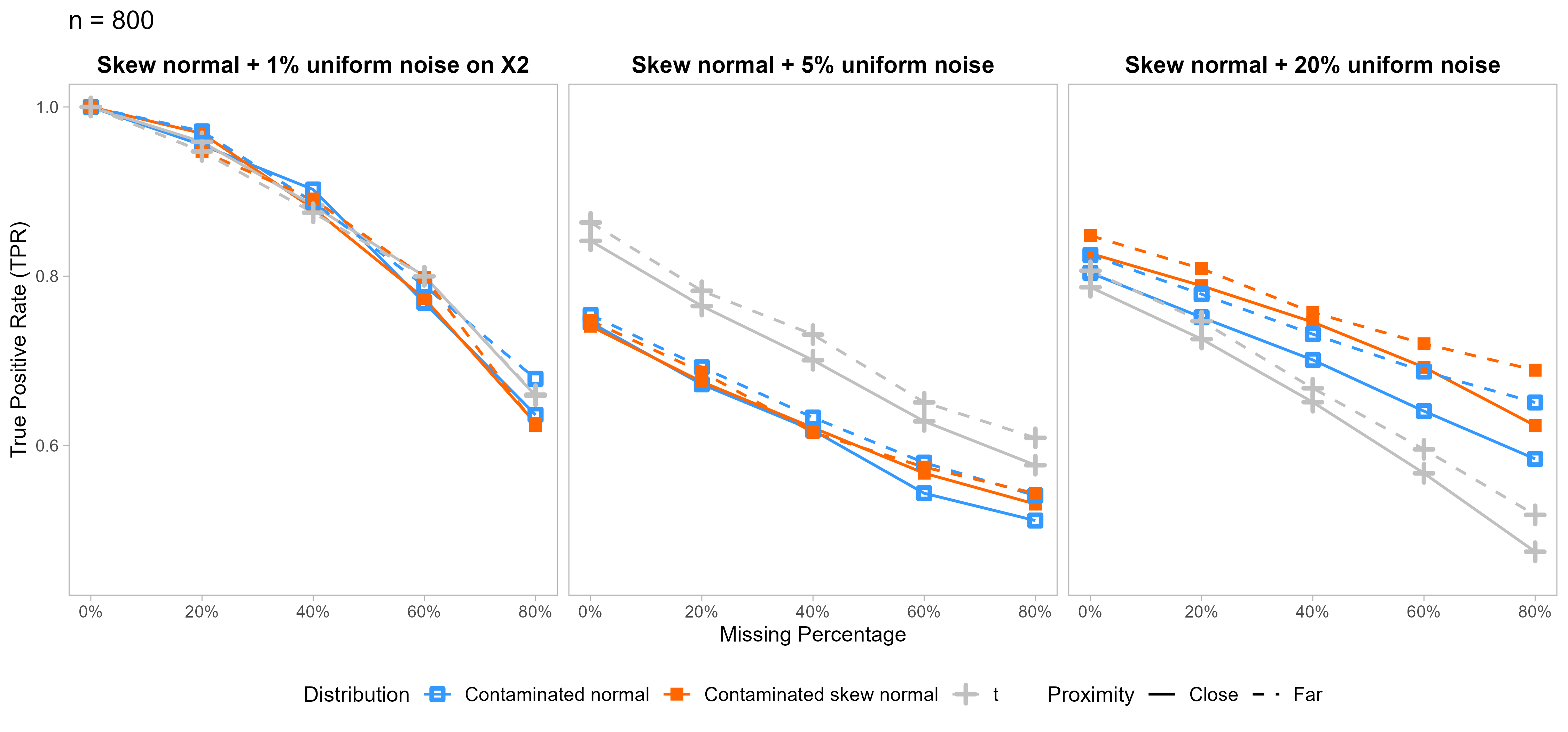}
    \caption{Average TPR values for $n=800$. Results are shown for the FMMt, FMCMN, and FMCMSN models across varying percentages of incomplete rows and across two levels of cluster proximity.}
    \label{tpr_800}
\end{figure}

\subsubsection{Simulation experiment: Part B}
The second part of the experiment concerns performance under high dimensionality. It constructs a dataset with one cluster sample of size $n=1000$ of dimension $p=10$. Effect of sample size is now excluded. The location and scale matrix are set to the zero vector and identity matrix, respectively. The skewness vector is kept at zero except at the tenth position. That is $\blam = [\bm{0}_9, \lambda_{10}]^{\top}$, where $\bm{0}_9$ is a vector of dimension $9$ filled with zeros, and $\lambda_{10} = 10$.

Datasets are generated under the following cases:
\begin{itemize}
    \item[(a)] FMMSN with 1\% of the observations of the sample are randomly selected and five of the observations are replaced by 0 and the other five are replaced with observations from a continuous uniform distribution on the interval (10, 15).
    \item[(b)] FMMSN with 5\% of rows randomly replaced by noise. The first nine entries $(X_1,\dots,X_9)$ are replaced with the same observed noise points on the interval $(-5,5)$ while the last entry $X_{10}$ observed from a continuous uniform distribution on the interval $(-5,5)$.
\end{itemize}
As in part A, the proportion of rows containing missing values is varied across 0\%, 20\%, 40\%, 60\%, and 80\% for each case. The accuracy rates, FPRs, and TPRs are all reported as well. 

\begin{figure}[H]
    \centering
    \begin{subfigure}{0.3\textwidth}
        \includegraphics[trim={0cm 0cm 0cm 2cm}, clip, width=0.9\textwidth]{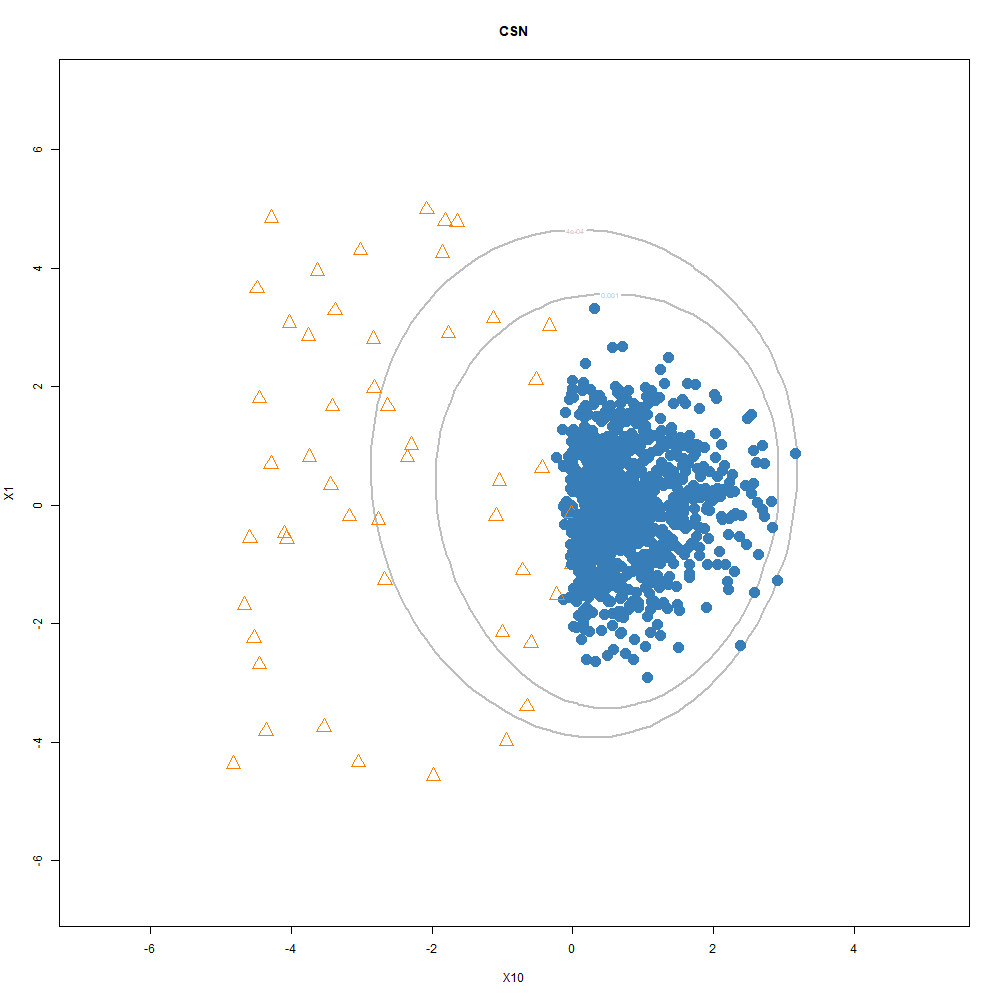}
        \caption{CMSN fitted model.}
    \end{subfigure}
    \begin{subfigure}{0.3\textwidth}
        \includegraphics[trim={0cm 0cm 0cm 2cm}, clip, width=0.9\textwidth]{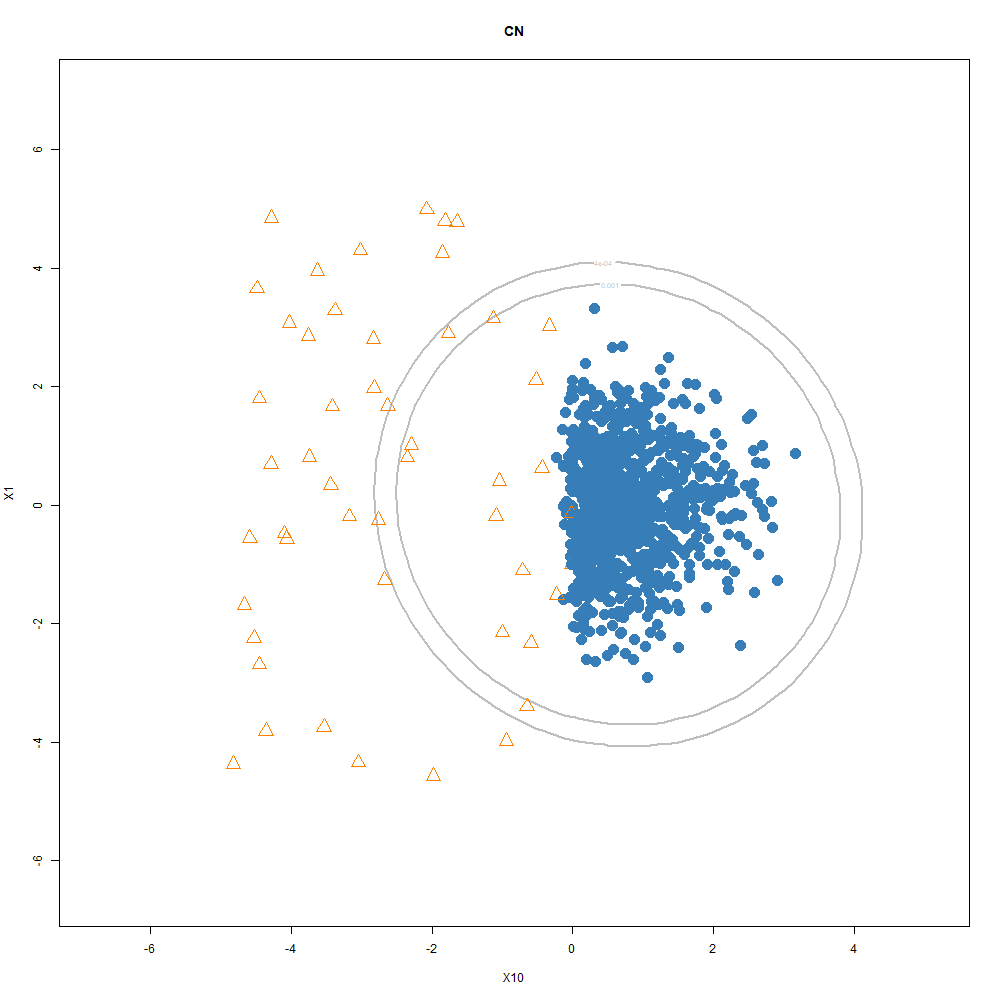}
        \caption{CMN fitted model.}
    \end{subfigure}
    \begin{subfigure}{0.3\textwidth}
        \includegraphics[trim={0cm 0cm 0cm 2cm}, clip, width=0.9\textwidth]{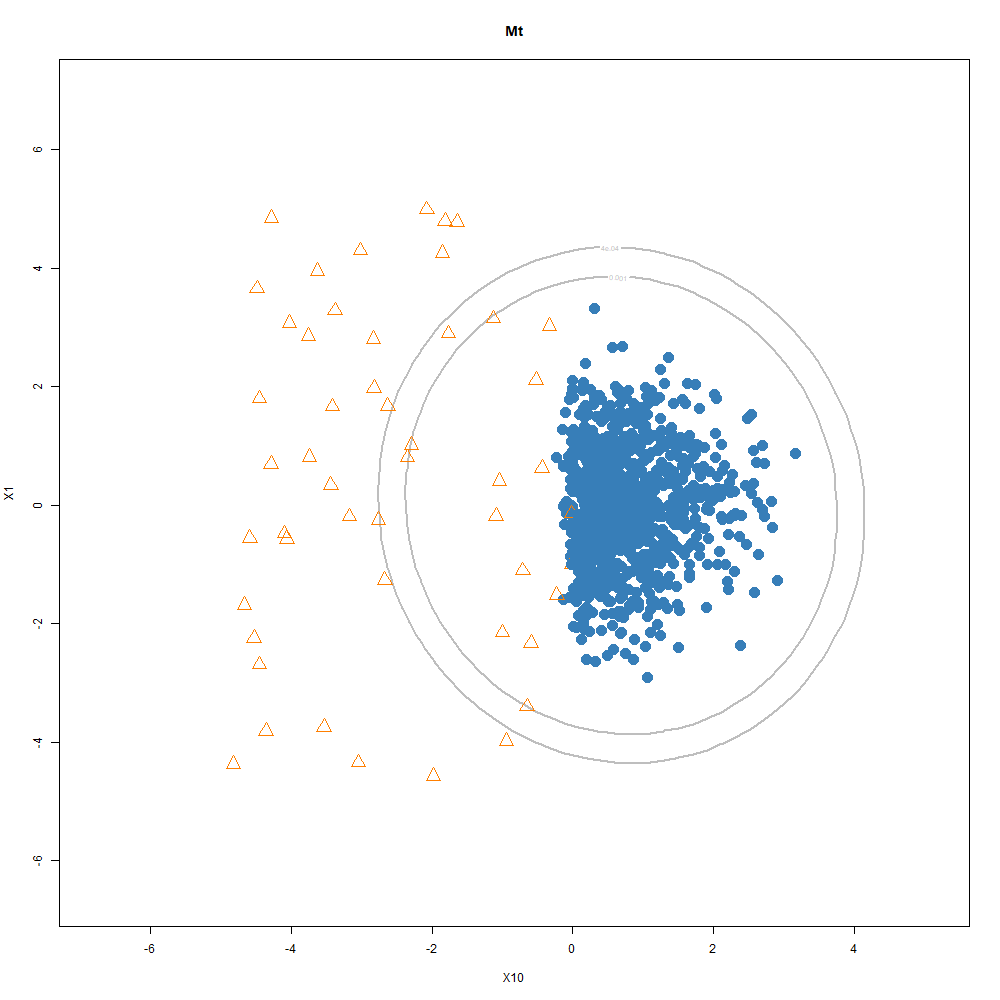}
        \caption{Mt fitted model.}
    \end{subfigure}
    \caption{Contour plot of the fitted models for variables $X_{1}$ against  $X_{10}$. The good and bad points are coloured in blue and orange respectively.}
    \label{contour_x10}
\end{figure}

\begin{figure}[H]
    \centering
        \begin{subfigure}{0.3\textwidth}
        \includegraphics[trim={0cm 0cm 0cm 2cm}, clip, width=0.9\textwidth]{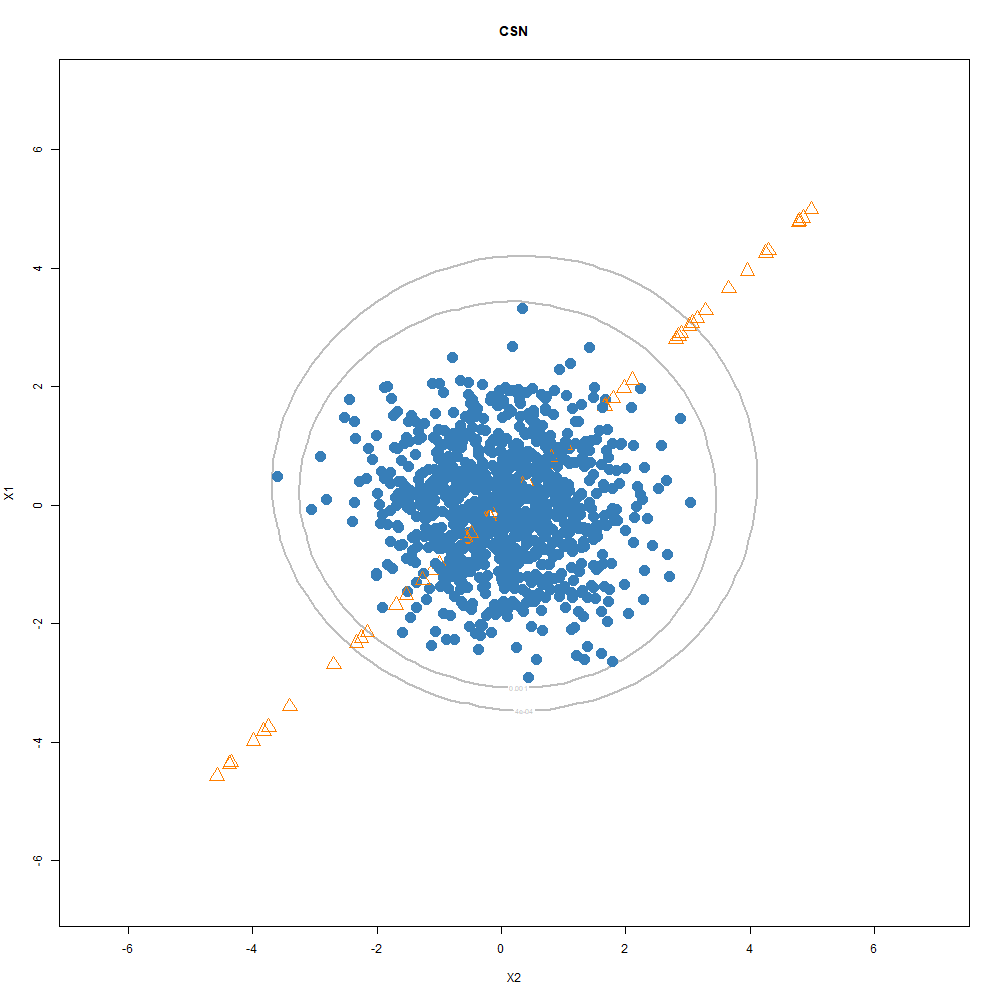}
        \caption{CMSN fitted model.}
    \end{subfigure}
    \begin{subfigure}{0.3\textwidth}
        \includegraphics[trim={0cm 0cm 0cm 2cm}, clip, width=0.9\textwidth]{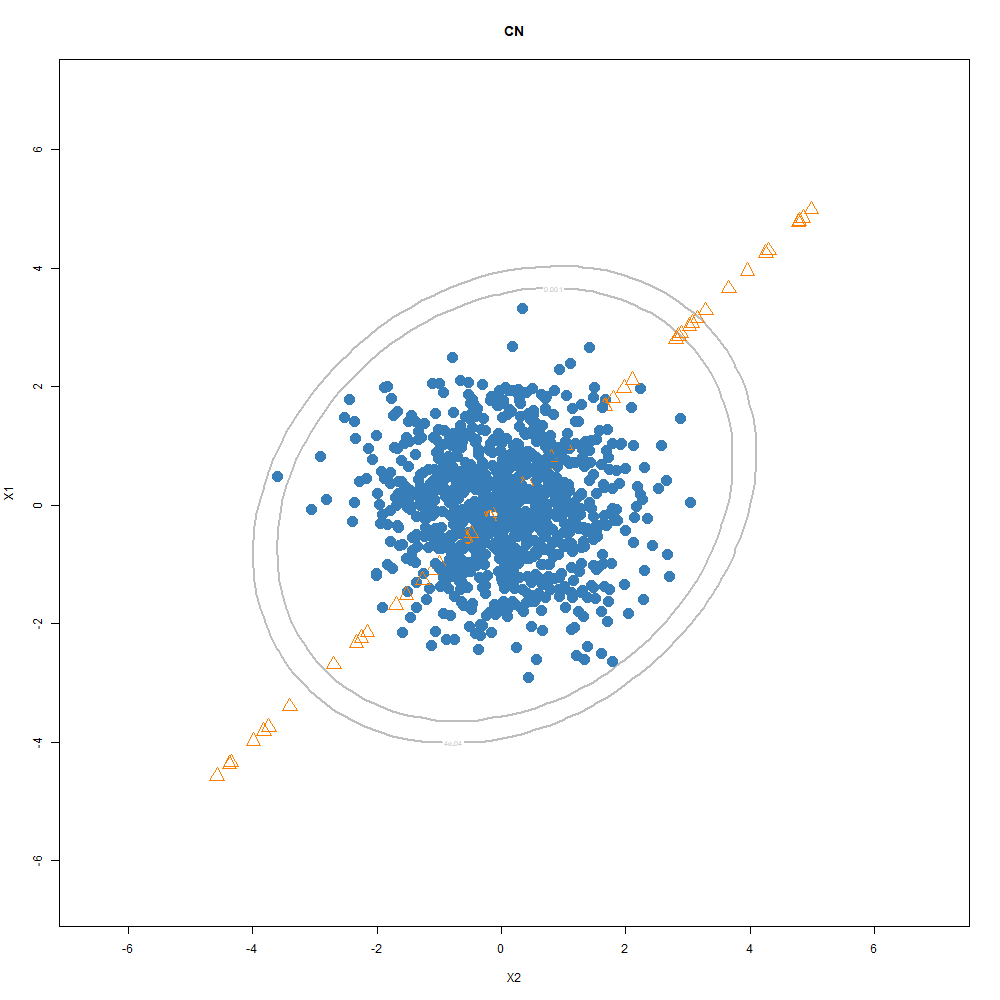}
        \caption{CMN fitted model.}
    \end{subfigure}
    \begin{subfigure}{0.3\textwidth}
        \includegraphics[trim={0cm 0cm 0cm 2cm}, clip, width=0.9\textwidth]{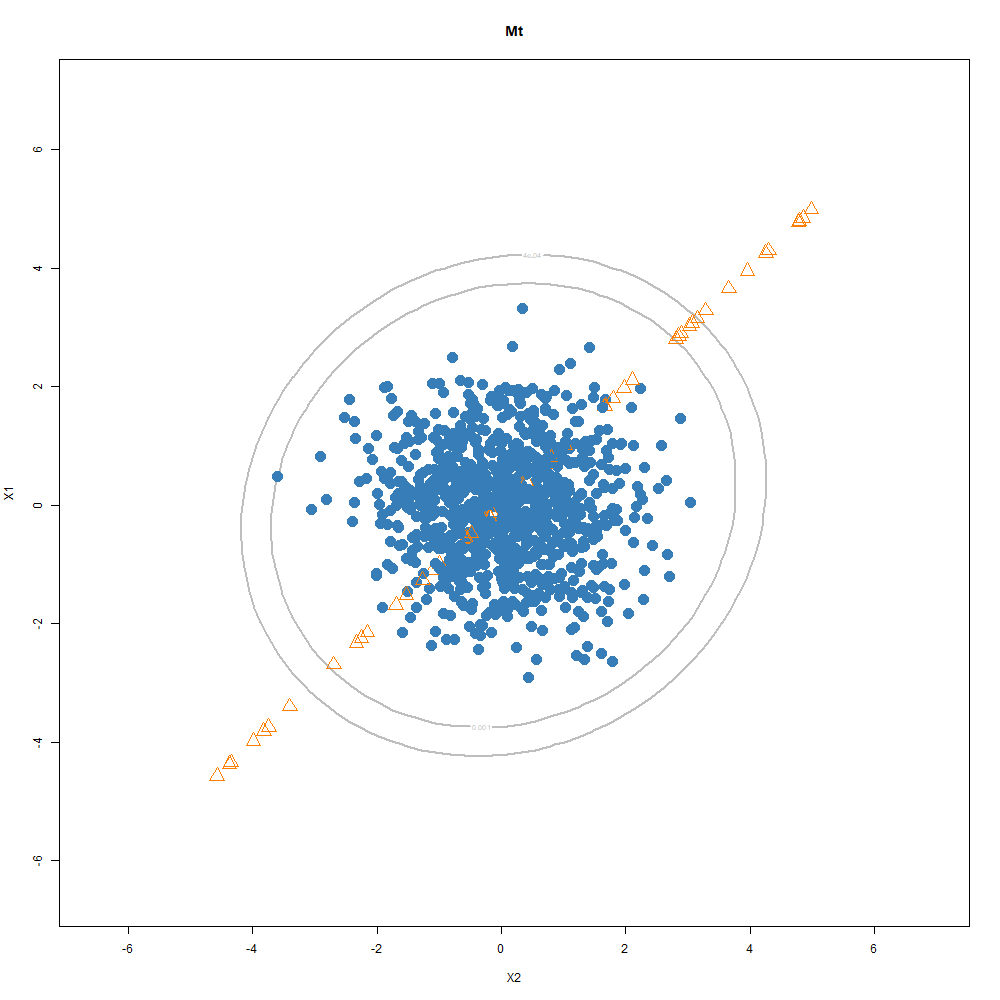}
        \caption{Mt fitted model.}
    \end{subfigure}
    \caption{Contour plot of the fitted models for variables $X_{1}$ against  $X_2$. The good and bad points are coloured in blue and orange respectively.}
    \label{contour_x1}
\end{figure}

\begin{figure}[H]
    \centering
    \includegraphics[trim={0cm 0cm 0cm 1cm}, clip, width=\textwidth]{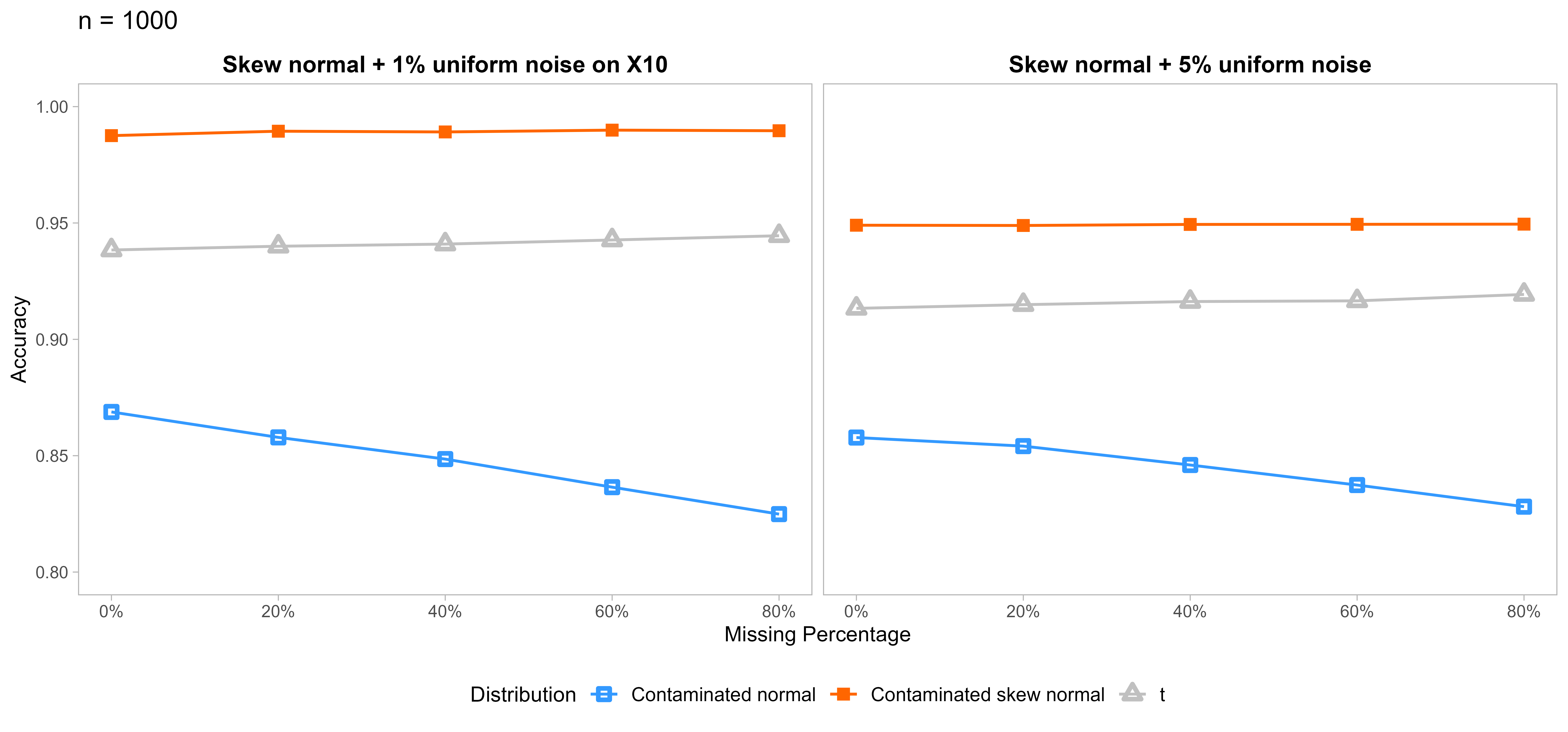}
    \caption{Average accuracy rates for $n=1000$. Results are shown for the FMMt, FMCMN, and FMCMSN models across varying percentages of incomplete rows.}
    \label{acc_1000}
\end{figure}
\begin{figure}[H]
    \centering
    \includegraphics[trim={0cm 0cm 0cm 1cm}, clip, width=\textwidth]{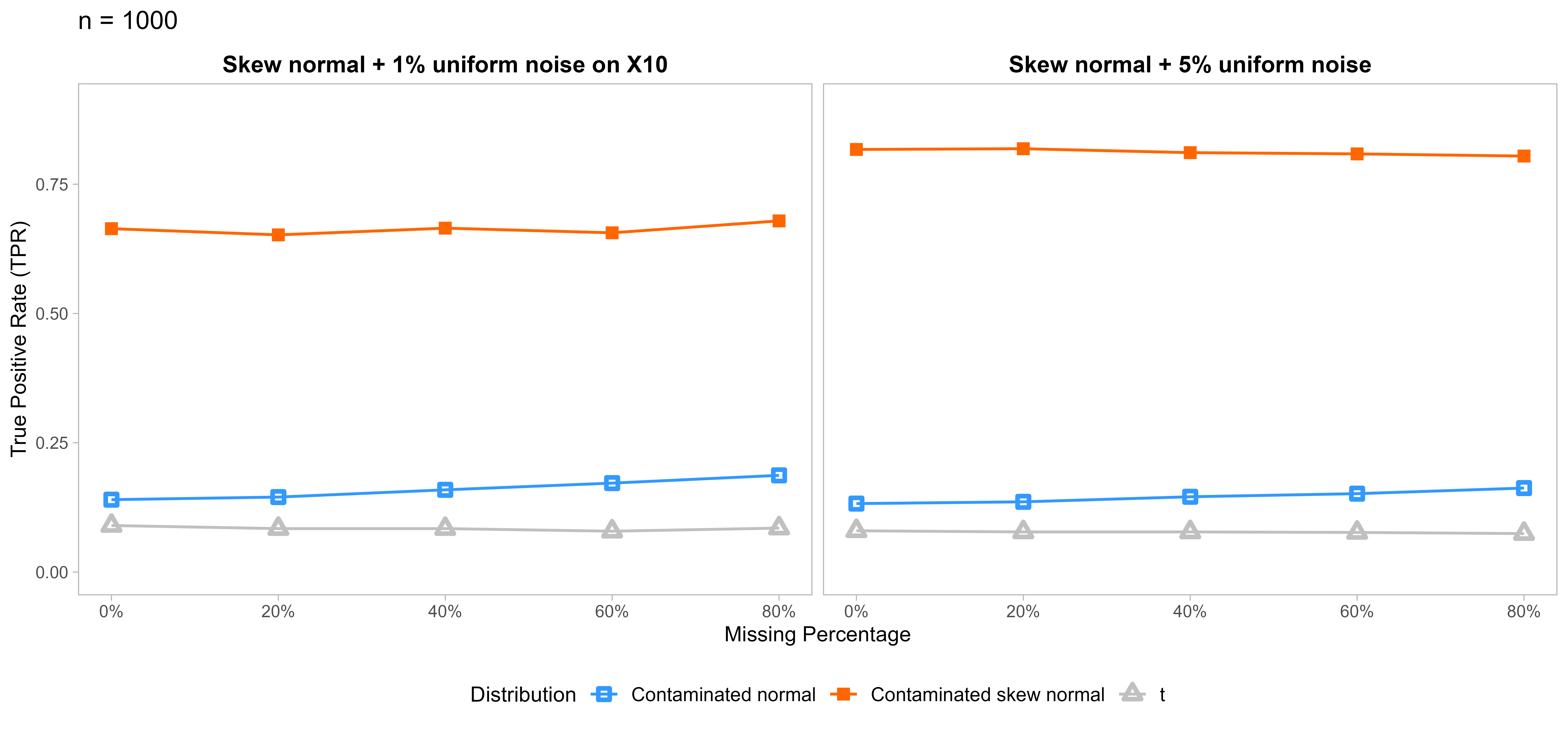}
    \caption{Average TPRs. Results are shown for the FMMt, FMCMN, and FMCMSN models across varying percentages of incomplete rows.}
    \label{tpr_1000}
\end{figure}
\begin{figure}[H]
    \centering
    \includegraphics[trim={0cm 0cm 0cm 1cm}, clip, width=\textwidth]{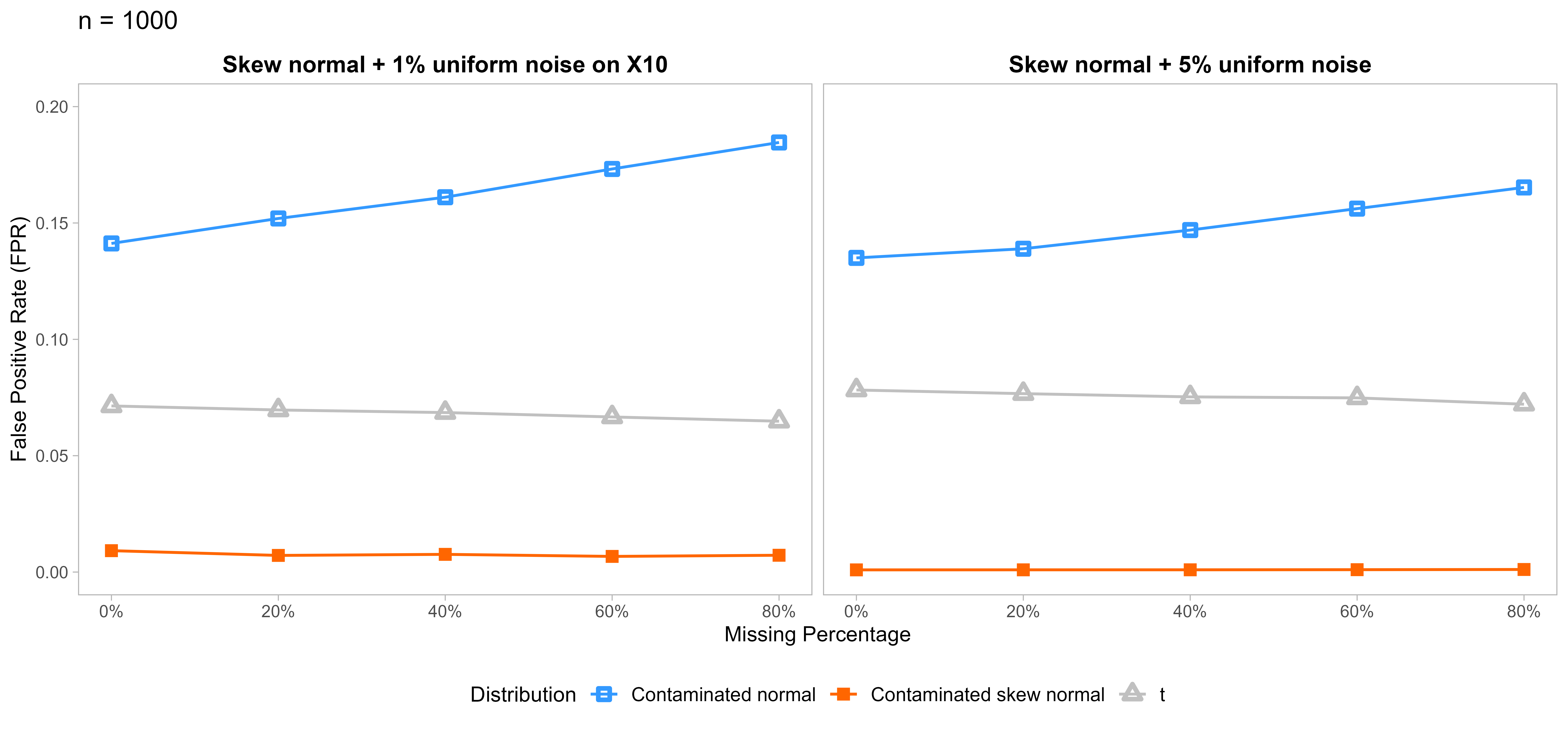}
    \caption{Average FPRs. Results are shown for the FMMt, FMCMN, and FMCMSN models across varying percentages of incomplete rows.}
    \label{fpr_1000}
\end{figure}

Looking at Figures \ref{acc_1000}, \ref{tpr_1000}, and \ref{fpr_1000}, there is a clear distinction in performance across the models for all three rates. 
The CMSN is the best performer across all metrics, demonstrating its ability to distinguish between good and bad points. 
A possible explanation for the observed trends can be inferred by comparing contour plots from a single experiment run in which 5\% of the data points were replaced with bad points, as described in scenario (b), shown in Figures \ref{contour_x10} and \ref{contour_x1}. 
As shown in \figurename~\ref{contour_x10}, the CMSN pdf has heavier tails in the direction of the outliers that accommodate the bad points (on the left). 
By design, for variables $X_1$ through $X_9$, the data distribution is not skewed, and the scale matrix is set to the identity. 
In \figurename~\ref{contour_x1}, we would expect the Mt and CMN contours to be circular, but the outliers have altered their shapes to ellipses, with the CMN most affected. 
The CMSN appears least affected in this regard, although the outliers have slightly affected its skewness. 

\section{Data application}
\label{application}
We apply the methodology to the Cleveland Children's Sleep and Health Study (CCSHS) dataset, made available by the National Sleep Research Resource (NSRR) repository. 
CCSHS is one of the largest population-based paediatric cohorts studied with objective sleep studies. The cohort is a stratified random sample of full-term (FT) and preterm (PT) children, born between 1988 and 1993, identified from the birth records of 3 Cleveland area hospitals (\cite{ccshs, cchs2}). The raw dataset consists of 255 variables and 517 observations, with a mix of categorical and continuous variables. The data can be requested from the NSRR website via application form fill-out. The dataset comes with a variable-dictionary, consisting of a column that specifies whether a variable is commonly used or not. Since the dataset contains multiple variables that represent decomposed components of the same underlying concept (e.g., fat, carbohydrates, protein, etc. as grams of nutrition), it is more parsimonious to use the aggregated total column rather than the individual components where applicable.

A quick overview of the subsetted dataset reveals a significant proportion of missing values, as shown in \tablename ~\ref{overview}.
   \begin{table}[H]
   \centering
    \caption{Proportion of missing values per variable from the CCSHS dataset.}
    \label{overview}
    \renewcommand{\arraystretch}{1}
   \begin{tabular}{llS}
   \toprule
    Variable & Name                                          & {Missing (\% )}\\
    \midrule
    bpsys     & Systolic blood pressure (SBP) (mean of 6 measurements)                                           & 0.00  \\
    bydias    & Diastolic blood pressure (DBP) (mean of 6 measurements)                                          & 0.00  \\
    bmi       & Body Mass Index (BMI)                                                                            & 0.00  \\
    mslp      & Average daily total sleep duration in main sleep in all days from actigraphy                     &13.15  \\
    cslp      & Coefficient of variation of daily total sleep duration in main sleep in all days from actigraphy & 15.67 \\
    mseff     & Average daily sleep efficiency in all days from actigraphy                                       &13.15  \\
    mrigrams  & Mean total grams per day                                                                         & 0.00  \\
    pbmi\_mom & Body Mass Index (BMI) of subject's mother                                                       & 18.18 \\
    pbmi\_dad & Body Mass Index (BMI) of subject's father                                                       & 67.12 \\
    \bottomrule
       \end{tabular}
       \label{colwise missing}
   \end{table}



\begin{figure}[H]
    \centering
    \includegraphics[width=0.7\textwidth]{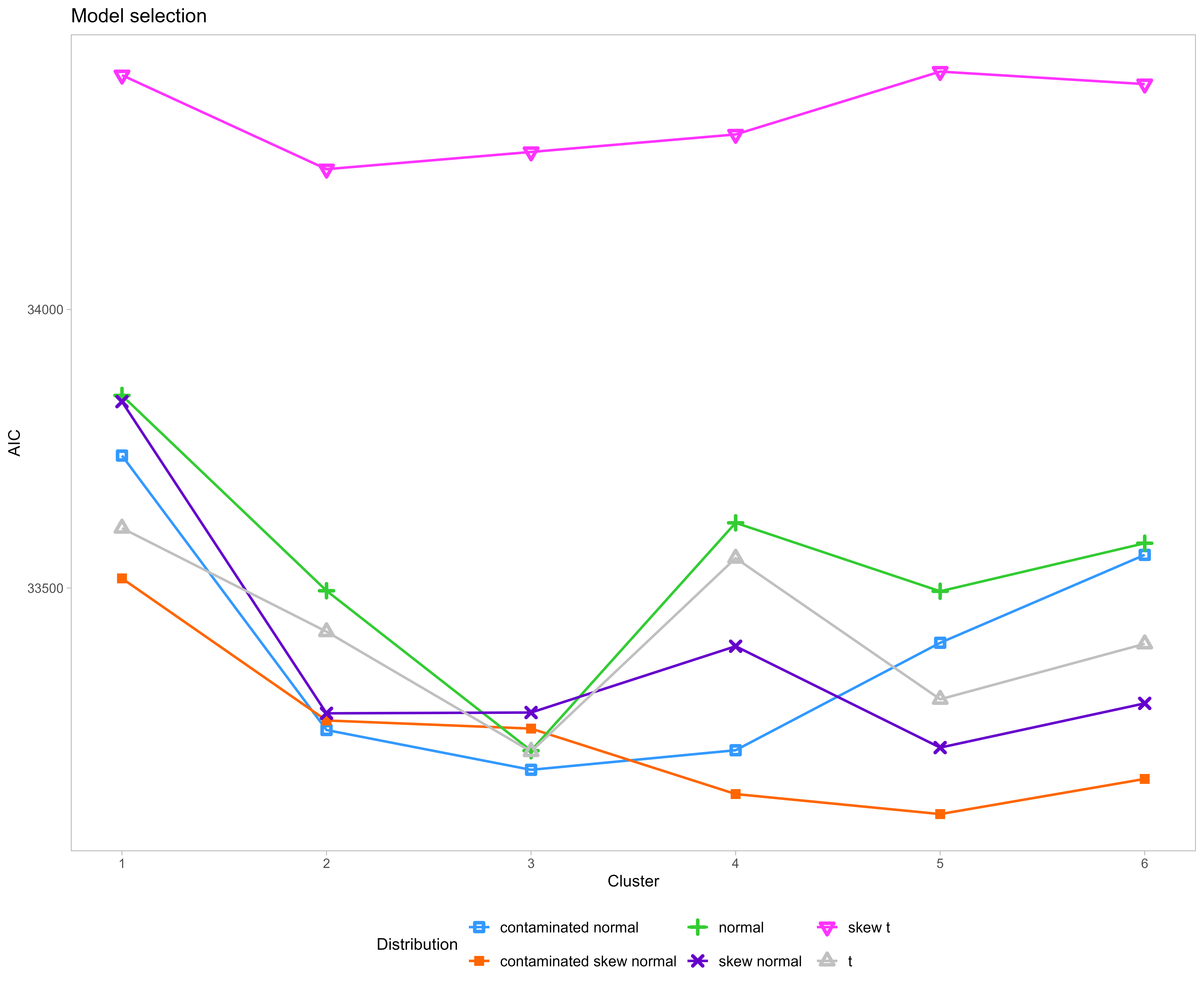}
    \caption{AIC values (vertical axis) vs the number of clusters chosen (horizontal axis) for the FMCSN model and its competitors. }
    \label{aic}
\end{figure}

We fit the proposed algorithm using several different numbers of clusters and compared it with other models using the AIC. The model with the lowest AIC was selected as the most appropriate fit. The variety of models under consideration remains within the realm of mixtures of contaminated and skewed distributions. It is not realistic to regard one of the candidates discussed in this paper as true generator behind the dataset. The model selection criteria should then reflect that the closest approximation to the true model behind the data's generation is chosen, rather than the true model itself. Thus, the AIC is a more practical metric for model selection (\cite{punzo2021multivariate, tortora2024laplace}).  From Figure \ref{aic} the FMCMSN model achieves the best performance for four, five, and six clusters, demonstrating strong overall competitiveness.

\begin{table}[H]
    \centering
    \caption{Some parameter estimates of the best fitting model: A 5-component CSN mixture model.}
    \label{csn_est}
        \renewcommand{\arraystretch}{1}
     \begin{tabular}{lSSSSS}
     \toprule
        Parameter   & \multicolumn{1}{c}{$g=1$}
                    & \multicolumn{1}{c}{$g=2$} 
                    & \multicolumn{1}{c}{$g=3$} 
                    & \multicolumn{1}{c}{$g=4$} 
                    & \multicolumn{1}{c}{$g=5$} \\
     \midrule
        $\hat{ \pi}_g$         & 0.249  & 0.116   & 0.186  & 0.338  & 0.111 \\
        $\hat{ \alpha}_g$      & 1.000  & 0.965   & 0.980  & 1.000  & 0.914 \\
        $\hat{ \beta}_g$       & 1.001  & 18.571  & 1.270  & 1.001  & 5.553 \\
        $\hat{ \lambda}_{1g}$  & 2.031  & 0.322   & 1.085  & 1.480  & 0.536  \\
        $\hat{ \lambda}_{2g}$  & 0.570  & 4.916   & 3.356  &  1.030 &  2.659 \\
        $\hat{ \lambda}_{3g}$  & 11.314 & 15.274  &  9.465 &  4.348 &  1.101 \\
        $\hat{ \lambda}_{4g}$  & -2.410 & -0.521  & -0.593 &  1.949 &  2.354 \\
        $\hat{ \lambda}_{5g}$  & -0.423 & -5.424  & -1.149 & -0.479 &  6.604 \\
        $\hat{ \lambda}_{6g}$  & -1.564 & -0.788  &  0.827 & -0.716 & -0.572 \\
        $\hat{ \lambda}_{7g}$  & -4.203 &  7.475  & -7.782 & -1.423 &  3.068 \\
        $\hat{ \lambda}_{8g}$  & -0.848 & -2.104  &  3.555 & -0.667 &  0.242 \\
        $\hat{ \lambda}_{9g}$  & 0.022  & -2.122  &  1.709 &  3.405 & -2.325 \\
        \bottomrule
    \end{tabular}
\end{table}

\begin{figure}[H]
    \centering
    \includegraphics[width=0.7\linewidth]{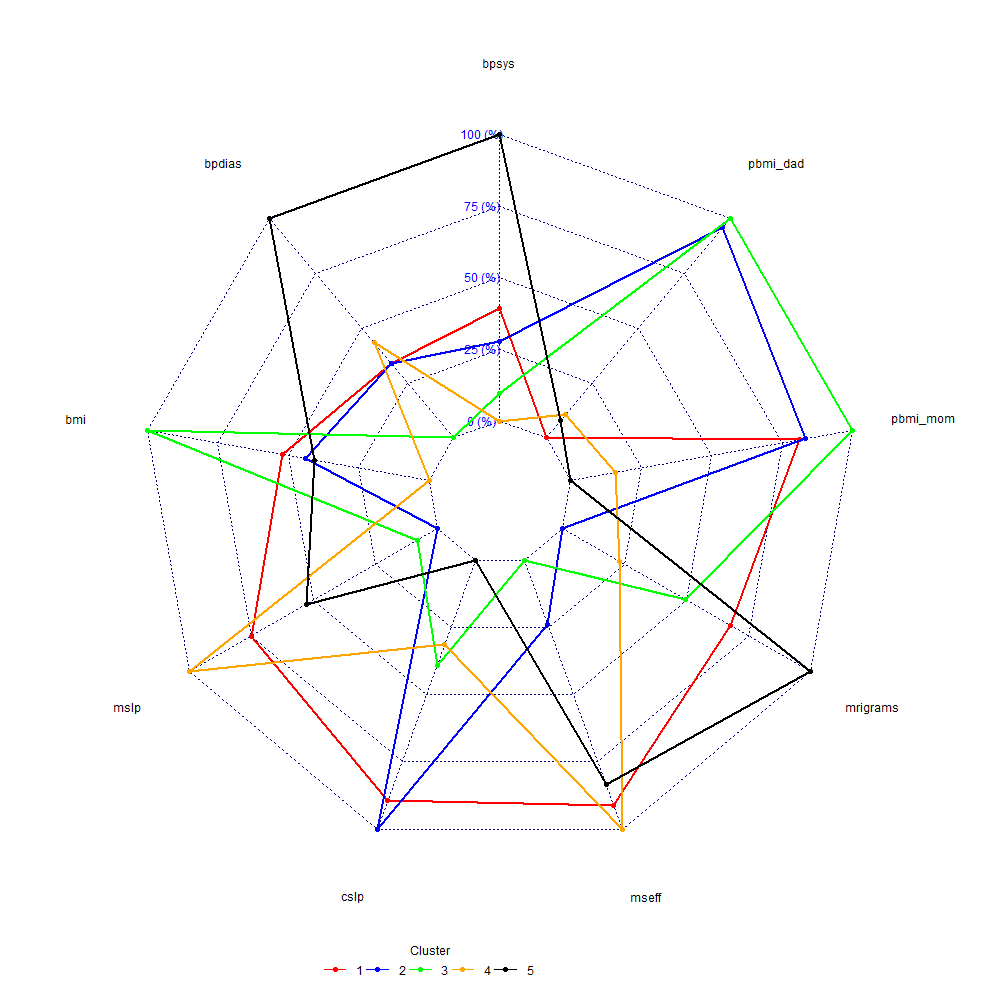}
    \caption{Radar plot of each cluster's medians for each variable expressed as percentages of the largest median.}
    \label{radarplot}
\end{figure}

Following the mixture model that minimises the AIC, we examine the results of a 5-component FMCMSN mixture model. Figure \ref{radarplot} presents the median of each variable used in the clustering fit for each cluster. We interpret these clusters using Figure \ref{radarplot} and a cluster-wise summary of some of the variables that were not used in the model fitting process to gain some further insights, provided in tables \ref{clsmean} and \ref{sleep}. The data can be segmented as follows:
\begin{itemize}
    \item Cluster 1: Delayed REM Sleepers\newline
          Individuals exhibit large mean daily sleep and high sleep efficiency, although sleep duration shows substantial variability. Both maternal BMI and mean daily food intake are elevated. Individuals in this cohort required the longest average time to enter REM sleep and the second-longest time from lights-out to sleep onset. The cluster is composed of over 56\% white and more than 41\% black participants, with 60\% male. This cluster demonstrates high-quality sleep, but individuals experience prolonged sleep onset and delayed entry into REM sleep.
    \item Cluster 2: Low Nutritional, Interrupted Sleepers \newline
         The lowest sleep efficiency lowest in average daily sleep duration, with high variability in sleep length is a key feature of this cluster. This cohort may have lower metabolic rates, as indicated by the lowest daily nutritional intake among all clusters, despite not having the lowest BMIs. The cluster has the largest proportion of non-white participants (over 56\% people of colour, 50\% black) and consists predominantly of females ($\approx$ 60\%). Delayed sleep onset is also characteristic. Sleep in this cluster is fragmented, with delayed onset and REM sleep, likely linked to low nutritional intake.
    \item Cluster 3: High BMI, Short-Duration Sleepers\newline
          Individuals in Cluster 3 have the highest BMIs in the study, mirrored by elevated parental BMIs. Surprisingly, diastolic blood pressure is lower while systolic pressure is higher. Sleep efficiency is low, with shorter but consistent sleep durations. Onset and REM sleep are relatively fast, ranking second among all clusters. This cluster experiences short-duration sleep similar to Cluster 2 but without reduced nutritional intake. Accumulated sleep debt may accelerate sleep onset and REM timing.
    \item Cluster 4: Lower BMI, Efficient Sleepers\newline
          Cluster 4 exhibits the highest sleep efficiency, large average daily sleep, and low variability. Nutritional intake is lower than other clusters, and both patient and parental BMIs are among the lowest. Diastolic blood pressure is higher than other clusters, with the lowest systolic readings typical of children and adolescents. The cluster comprises 65\% females and 71\% white participants. Onset sleep is rapid, though time to REM sleep is longer. Individuals in this cluster display generally healthy sleep patterns.
    \item Cluster 5: High Blood Pressure, High Nutritional Intake.\newline
          This cluster has high sleep efficiency, elevated blood pressure, and the second-highest BMI despite lower parental BMIs. Predominantly male (82\%), these individuals consume higher daily nutritional and fat intakes, with substantial snack and carbohydrate consumption. Time to sleep onset is longest, yet time from onset to REM sleep is the shortest. Prolonged sleep onset may reflect insomnia or anxiety, with rapid REM onset indicating the body’s attempt to compensate.
    \item Contamination\newline
          Clusters 1 and 4 comprise a proportion of good points estimated as 1, as given by table \tablename~\ref{csn_est}. These two clusters, observed to display high sleep efficiency and a generally higher level of sleep quality, do not present contamination, which is to be expected. Contamination is present in clusters 2, 3, and 5. Cluster 2 displays the largest degree of contamination, with an estimation degree of contamination around 18.571. where high variability in sleep length and sleep efficiency are the main contributors. Cluster 5 has the largest estimated proportion of bad points, making up 8.6\% of the cluster's size. 
\end{itemize}

Overall, this dataset shows the kinds of problems that come up when analysing data in sleep research and, more broadly, in medical studies. Using only the complete cases would be unwise, since it would keep less than 33\% of the rows. The estimated skewness vectors in Table \tablename~\ref{csn_est} suggest that the identified clusters are skewed. further indicate that the identified clusters are skewed. When a mixture model that assumes cluster-wise symmetry is applied, the number of clusters selected becomes sensitive to the choice of selection criterion. In contrast, the FMCMSN model accommodates the leptokurtosis present in each cluster and provides an interpretable framework for it, a feature not shared by other skewed distributions. This data application reflects a recurring pattern in medical datasets. Characteristics such as skewness, heavy tails, and, most notably, missing values may carry useful information. Highlighting this perspective supports the use of the algorithm that fits the FMCMSN model.

\begin{table}[H]
\centering
\caption{Cluster-wise average of variables used in the cluster algorithm.}
\label{clsmean}
    \renewcommand{\arraystretch}{1}
\begin{tabular}{l
S[table-format=3.5]
S[table-format=3.5]
S[table-format=3.5]
S[table-format=3.5]
S[table-format=3.5]}
\toprule
Variable & {1} & {2} & {3} & {4} & {5} \\
\midrule
Systolic blood pressure (SBP) (mean of 6 measurements)       & 116.30450 & 115.38280 & 114.84510 & 111.85330 & 124.09940 \\
Diastolic blood pressure (DBP) (mean of 6 measurements)      & 63.36745  & 65.37760  & 62.53333  & 64.29620  & 69.92982  \\
Body Mass Index (BMI)                                        & 25.79952  & 25.68412  & 28.45436  & 22.58288  & 25.85427  \\
Average daily total sleep duration                           & 459.10260 & 419.12210 & 430.39840 & 474.18860 & 462.73240 \\
Coefficient of variation of daily total sleep duration       & 21.33427  & 22.91350  & 16.42136  & 15.22711  & 13.70629  \\
Average daily sleep efficiency in all days from actigraphy   & 96.61690  & 90.84843  & 91.51660  & 97.25323  & 96.61840  \\
Mean total grams per day                                     & 2400.530  & 1370.692  & 2218.187  & 1702.336  & 3248.985  \\
Body Mass Index (BMI) of subject's mother                    & 31.54915  & 33.67910  & 35.82103  & 26.59587  & 27.59972  \\
Body Mass Index (BMI) of subject's father                    & 28.01292  & 33.92614  & 33.85710  & 28.85481  & 31.00806  \\
\bottomrule
\end{tabular}
\label{cluster_summary}
\end{table}

\begin{table}[H]
\centering
\caption{Cluster-wise average of sleep variables from the dataset.}
\label{sleep}
    \renewcommand{\arraystretch}{1}
\begin{tabular}{l
S[table-format=2.5]
S[table-format=2.5]
S[table-format=2.5]
S[table-format=2.5]
S[table-format=2.5]}
\toprule
Variable & {1} & {2} & {3} & {4} & {5} \\
\midrule
Sleep maintenance efficiency  & 25.21260 & 26.34375 & 22.36471 & 17.14835 & 31.22807 \\
REM sleep latency including wake from type I polysomnography   & 128.25980 & 123.84380 & 118.70590 & 123.25820 & 116.87720 \\
REM sleep latency excluding wake from type I polysomnography  & 113.18110 & 106.75000 & 103.97647 & 110.15934 & 98.52632 \\
Total Sleep Duration from type I polysomnography   & 460.17320 & 460.17190 & 459.75290 & 482.09890 & 443.73680 \\
\bottomrule
\end{tabular}
\label{cluster_sleep_summary}
\end{table}

\section{Conclusion}
\label{conclusion}
Data collected from sleep studies—and medical research more broadly—is inherently heterogeneous, often exhibiting skewness, outliers, and missing values. A suitable model to capture this heterogeneity must therefore reflect the data’s complex statistical characteristics. This paper proposes a unified clustering algorithm designed to model such intricate data structures. Specifically, it extends the finite mixture of contaminated multivariate skew-normal (FMCMSN) distributions—which already accounts for skewed clusters and within-cluster outliers—to simultaneously accommodate incomplete data. This represents a substantial improvement over conventional approaches that treat missing values as a separate preprocessing step.
The principal contribution of this work lies in modifying the FMCMSN estimation procedure to jointly address missingness and contamination within a single modelling framework. The contamination parameters, $\alpha_g$ and $\beta_g$, allow the mixture model to automatically account for anomalous observations while detecting outliers intrinsically, avoiding the need for \textit{ad hoc} identification procedures that disregard underlying statistical structure. Simulation studies demonstrate that the proposed algorithm performs competitively in clustering accuracy—comparable to the finite mixture of multivariate skew-$t$ (FMMSt) model—while offering the additional advantage of automatic outlier detection.
Beyond its methodological benefits, the proposed model provides enhanced interpretability in the context of sleep research. It facilitates the identification of patterns and supports robust population-level inference without requiring prior data cleaning or the exclusion of incomplete cases.
Applied to the Cleveland Children’s Sleep and Health Study (CCSHS) dataset, the method identifies five distinct groups of sleepers—Delayed REM Sleepers, Low-Nutritional Interrupted Sleepers, High-BMI Short-Duration Sleepers, Lower-BMI Efficient Sleepers, and High-Blood-Pressure High-Nutritional-Intake Sleepers—highlighting important differences in sleeper typologies. Moreover, outliers were automatically detected.
Overall, this work underscores the importance of statistical methodologies capable of capturing skewed clusters, detecting outliers, and handling missingness as intrinsic features of real-world biomedical data.

\section{Acknowledgements}
The Cleveland Children's Sleep and Health Study (CCSHS) was supported by grants from the National Institutes of Health (RO1HL60957, K23 HL04426, RO1 NR02707, M01 Rrmpd0380-39). The National Sleep Research Resource was supported by the National Heart, Lung, and Blood Institute (R24 HL114473, 75N92019R002).

This work is supported in part by the Centre of Excellence in Mathematical and Statistical Sciences, 363 based at the University of the Witwatersrand (SA), grant number PMDS230705128094 as well as the Department of Research and Innovation (DRI).The opinions expressed and conclusions arrived at are those 364 of the authors and are not necessarily to be attributed to the NRF.

Antonio Punzo acknowledges the support by the Italian Ministry of University and Research (MUR) under the PRIN 2022 grant number 2022XRHT8R (CUP: E53D23005950006), as part of “The SMILE Project: Statistical Modelling and Inference to Live the Environment”, funded by the European Union – Next Generation EU.

\appendix
\section{Supplementary material}
The derivations behind the expectations in the E step in section \ref{e_step} are now given. Firstly, Bayes theorem is used to determine that 
\begin{align}
\label{t_given_v1}
    T_i|\bXo_i = \bxo_i, V_{ig} = 1 ,Z_{ig} = 1 \sim TN(\mu^{(k)}_{T_{ig}}, \sigma^{2 (k)}_{T_{g}}),
\end{align}
and
\begin{align}
\label{t_given_v0}
    T_i|\bXo_i = \bxo_i, V_{ig} = 0,Z_{ig} = 1 \sim TN(\beta_g^{{(k)~ -1/2}}\mu^{(k)}_{T_{ig}}, \sigma^{2 (k)}_{T_{g}}),
\end{align}
where $\sigma^{2 (k)}_{T_{g}} = \left(1 + \bDelta_{o,g}^{{(k)}\top}(\bOmega_{oo,g}^{(k)})^{-1}\bDelta_{o,g}^{(k)} \right)^{-1}$ and $ \mu^{(k)}_{T_{ig}} = \sigma^{2 (k)}_{T_{g}} \bDelta_{o,g}^{{(k)}\top}(\bOmega_{oo,g}^{(k)})^{-1}(\bxo_i - \bmu_{o,g}^{(k)})$. Let $ W_{\phi}(\cdot) = \frac{\phi_1(\cdot)}{\Phi_1(\cdot)}$ and $A_{ig}^{o~(k)} =  (\dot{\blam}_{o,g}^{(k)})^{\top}(\bSig_{oo,g}^{(k)})^{-1/2}(\bxo_{i} - \bmu^{(k)}_{o,g})$. Using the moments from the distributions of \eqref{t_given_v1} and \eqref{t_given_v0} we obtain:
\begin{align}
    \vkt    &= \mathbb{E}\left[V_{i}T_i| Z_{i,g}, \bx^{o}_i   \right] \nonumber\\
            &= \mathbb{E}\left[V_{i} \mathbb{E}[T_i| Z_{i,g}, V_i=1,\bx^{o}_i ] \bx^{o}_i, Z_{i,g} = 1\right] \nonumber\\
            &=  \vk \left[\mu^{(k)}_{T_{ig}} + \sigma^{(k)}_{T_g}W\left( A^{o~(k)}_{ig} \right) \right],\\
    t_{ig}^{(k)}-\vkt   &= \mathbb{E}\left[(1-V_{i})T_i| Z_{i,g}, \bx^{o}_i   \right] \nonumber\\
                        &=  \mathbb{E}\left[(1-V_{i} )\mathbb{E}[T_i| Z_{i,g}, V_i=0,\bx^{o}_i ] \bx^{o}_i, Z_{i,g} = 1\right] \nonumber \\
                        &= (1-\vk)\left[ \beta_g^{{(k) ~ -\half}}\mu^{(k)}_{T_{ig}} + \sigma^{(k)}_{T_g}W_{\phi}\left( \beta_g^{(k) ~ -\half}A^{o~ (k)}_{ig} \right) \right],
\end{align}
and
\begin{align}
    \etwo   &= \mathbb{E}\left[V_{i}T_i^2 + (1-V_{i})T_i^2 | Z_{i,g}, \bx^{o}_i   \right]\nonumber\\
            &= \left[ \vk + \frac{1 - \vk}{\beta_g^{(k)} } \right]\mu_{T_{i,g}}^2 
                  + 2\left[\heta +  \frac{\hetac}{\beta_g^{(k) ~\half} }  \right]\mu^{(k)}_{T_{ig}}\sigma_{T_{g}} + \sigma^{2 (k)}_{T_{g}},
\end{align}
with  $\heta=  \vk W_{\phi}\left(A_{ig}^{o~(k)}\right)$ and $\hetac = (1 - \vk)W_{\phi}\left(\beta_g^{(k)-\half} A_{ig}^{o~(k)} \right)$.

The expectations of $\vx, \vcxx, \vxx$, and $\vcxk$ are computed using Theorem \ref{sn dists}, which produces:
\begin{align}
    \vx     & = \mathbb{E}\left[V_{i} \bXm_i |Z_{i,g}, \bx^{o}_i  \right]  = \vk\bmu^{(k)}_{c,g} + \heta \bDelta^{(k)}_{c,g},\\
    \vcxk   & =\mathbb{E}\left[(1-V_{i}) \bXm_i  |Z_{i,g}, \bx^{o}_i  \right] = (1-\vk)\beta_g^{(k) ~ -1}\bmu^{(k)}_{c,g} + \beta_g^{(k) -\half}\hetac\bDelta^{(k)}_{c,g},\\
     \vxx  & = \mathbb{E}\left[V_i\bXm_i\bX_i^{m \top} |Z_{i,g}, \bx^{o}_i  \right] 
              = \vk \bSig^{(k)}_{c,g} + \vk\bmu_{c,g}^{(k)}(\bmu_{c,g}^{(k)})^{\top} +\heta \bm{\xi}^{(k)}_{ig}, \\
    \label{vcxx}
    \vcxx  & = \mathbb{E}\left[(1-V_i)\bXm_i\bX_i^{m \top} |Z_{i,g}, \bx^{o}_i  \right] 
              = (1-\vk)\beta_g^{(k)} \bSig^{(k)}_{c,g} + (1-\vk)\bmu_{c,g}^{(k)}(\bmu_{c,g}^{(k)})^{\top}  +\hetac \bm{\xi}^{(k)}_{ig},
\end{align}
where $\bmu_{c,g}$ and $\bSig_{c,g}$ are given as in Theorem \ref{sn dists} and $\bm{\xi}^{(k)}_{i,g} = \bmu^{(k)}_{c,g}(\bDelta^{(k)}_{c,g})^{\top} + \bDelta^{(k)}_{c,g}(\bmu_{c,g}^{(k)})^{\top} -\bDelta_{c,g}^{(k)} (\bDelta_{c,g}^{(k)})^{\top}$. Using Theorem \ref{norm cond dist}, the following expected values are computed as:
\begin{align}
\vktx   & = \mathbb{E}\left[ V_iT_i\bX_i |Z_{i,g}, \bxo_i  \right] = \vkt\bm{m}^{(k)}_{c,g} + \vktt \bm{\gamma}^{(k)}_{c,g}, \\
\label{vctx}
\vctxk  & = \mathbb{E}\left[ (1-V_i) T_i\bX_i |Z_{i,g}, \bxo_i  \right] = \vkt\bm{m}^{(k)}_{c,g} + \vktt \beta_g^{(k) \half}\bm{\gamma}^{(k)}_{c,g}, 
\end{align}
where $ \bDelta^{(k)}_{c,g} =\frac{\bSig^{(k)1/2}_{c,g}  \blam^{(k)}_{c,g}}{ \sqrt{1 +  (\blam^{(k)}_{c,g})^{\top} \blam^{(k)}_{c,g} }}$.

\printbibliography

\end{document}